\documentclass[pra,aps, twocolumn,showpacs]{revtex4}
\usepackage{amsmath}
\usepackage{amssymb}
\usepackage{epsfig}
\pdfoutput=1

\begin{document}

\title{Measuring complete quantum states with a single observable}

\author{Xinhua Peng$^{1}$}
\email{xinhua@e3.physik.uni-dortmund.de}
\author{Jiangfeng Du$^{1,2}$}
\author{Dieter Suter$^{1}$}

\affiliation{$^{1}$Fachbereich Physik, Universit\"{a}t Dortmund,
44221 Dortmund, Germany}
\affiliation{$^{2}$Hefei National
Laboratory for Physical Sciences at Microscale and Department of
Modern Physics, University of Science and Technology of China,
Hefei, Anhui 230026, P.R. China}

\date{\today}

\begin{abstract}
Experimental determination of an unknown quantum state usually requires
several incompatible measurements.
However, it is also possible to determine the full quantum state from a single,
repeated measurement.
For this purpose, the quantum system whose state is to be
determined is first coupled to a second quantum system (the ``assistant") in such a way
that part of the information in the quantum state is transferred to the assistant.
The actual measurement is then performed on the enlarged system including
the original system and the assistant.
We discuss in detail the requirements of this procedure and experimentally implement
it on a simple quantum system consisting of nuclear spins.
\end{abstract}


\maketitle

\section{INTRODUCTION}

Given the state of a quantum system, one is able to calculate
the results of any measurement performed on that system.
However, to determine the state from the results of measurements, 
one usually has to perform different measurements
that are not mutually compatible, using non-commuting observables.
This issue is sometimes referred to as the "Pauli problem", since
Pauli discussed it in 1933  \cite{Pauli:1933aa}
Since then, interest in this issue has continued, as it touches the fundamentals
of quantum mechanics \cite{Bohr:1935aa}
More recently, it was also found to be of practical importance in quantum communication
   \cite{HelstromBook:1976aa, Chefles:2000aa, Gill:2003aa} , e.g., in quantum cryptography and
quantum key distribution \cite{Bechmann-Pasquinucci:2000aa,Bechmann-Pasquinucci:2000ab}.

If we consider an ensemble \textbf{\emph{S}} of $N$-level systems,
its quantum state is described by a density matrix
$\hat{\rho}$ in $N$-dimensional Hilbert space, which requires
$N^2-1$ real parameters for its complete specification.
These parameters can be determined experimentally from the outcomes
of a series of different measurements on identically prepared ensembles.
Since quantum mechanical measurements with an observable $\hat{\Omega}$ 
with the spectral decomposition
$\hat{\Omega} =\sum_{\alpha=1} ^N \varpi_\alpha \hat{P}_\alpha $
generate at most $N-1$ independent
probabilities, at least
$(N^2-1)/(N-1)=N+1$ measurements with noncommuting
observables are required to fully determine the unknown state $\hat{\rho}$. 

Techniques for the reconstruction of the complete quantum state from a series
of measurements are commonly referred to as ``quantum state tomography"
\cite{Welsch:1999aa, DAriano:1997aa, LeonhardtBook:1997aa}. 
Different versions of such techniques have been proposed, with the goal
of obtaining the best possible information about the unknown state 
while using the smallest possible number of measurements.
Since the number of measurements must be at least $N+1$,
the task is thus to determine an optimal set of $N+1$ observables
(see, e.g.,\cite{Ivonovic:1981aa}).
A solution to this problem was given by Wootters and Fields \cite{Wootters:1989aa}:
They found that the observables should be chosen such that their basis states
are evenly distributed through Hilbert space ``mutually unbiased").
This choice assures that the data redundancy is minimized and the
information content is maximized.

As the simplest example, we consider the state of a spin 1/2.
Its density operator can be written in the form
$\hat{\rho}=\frac{1}{2}(\mathbf{1}+\vec{s}\cdot\vec{\hat{\sigma}})$,
where $\vec{\hat{\sigma}}$ are the Pauli operators and $\vec{s}$ is a 
dimensionless vector of length $\le 1$ that specifies the position
of the state in the Bloch sphere. 

\begin{figure}[tbh]
\begin{centering}
\includegraphics[width=0.8\columnwidth]{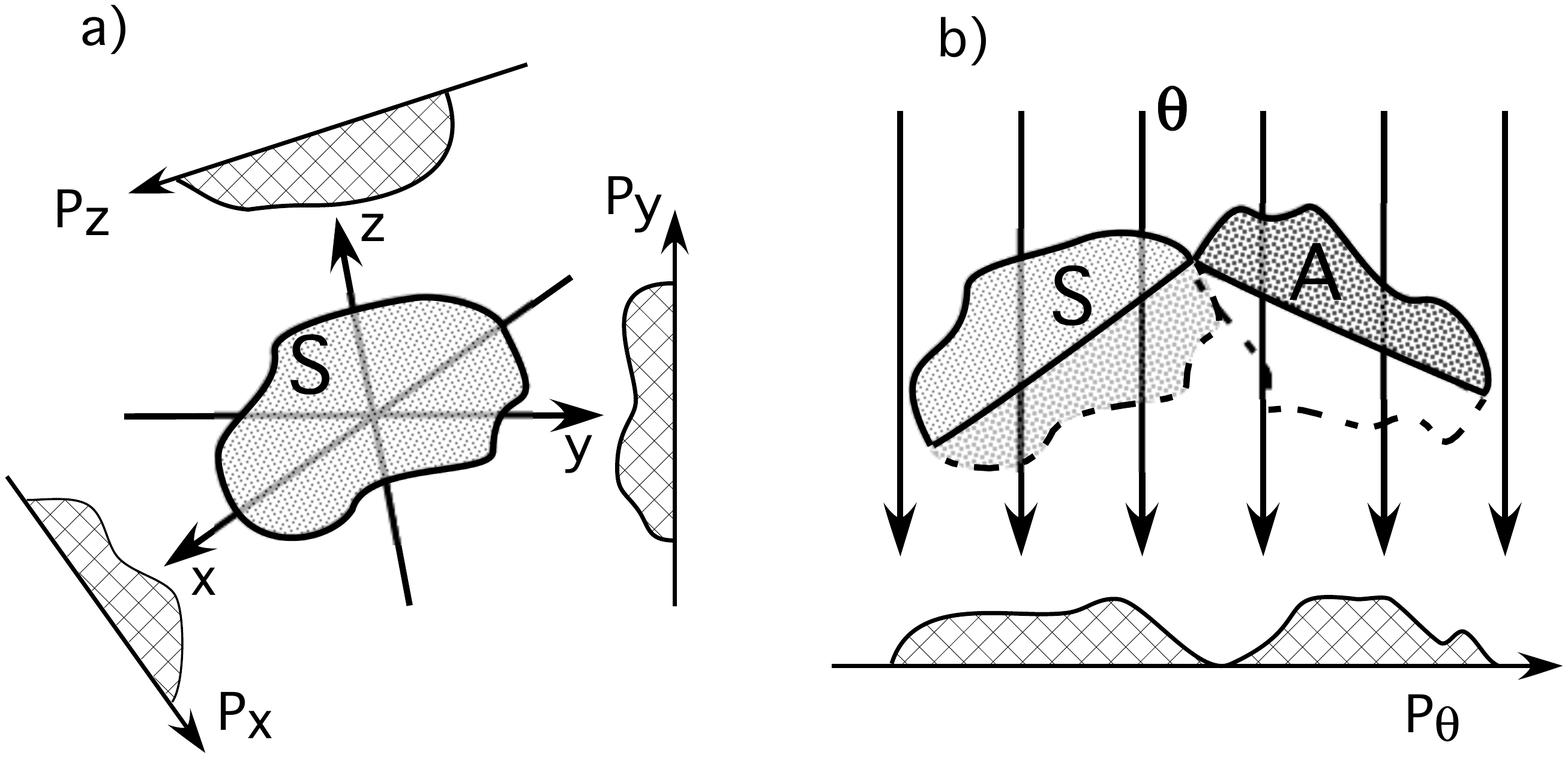}
\end{centering}
\caption{Schematic representation of different schemes for the experimental determination
of the state of a quantum system S.
(a) Quantum state tomography: a series of mutually incompatible measurements
are performed, projecting the quantum state e.g. along the x, y, and z-axes. 
(b) The present approach: Part of the information in the quantum state is first
transferred to the assistant A. A single measurement of the combined system S + A,
in direction $\theta$, can then determine the complete initial state of S.}
\label{SingleMeas}
\end{figure}

The simplest approach to determine this state 
consists of measuring the spin components along the 
$x, y$, and $z$-axes, yielding 6 possible measurement outcomes 
(see Fig. \ref{SingleMeas} a).
A minimal set of measurements only requires four such probabilities.
They may be chosen as the probabilities for measuring the spin operator 
components in four directions that are oriented like the face normals 
of a tetrahedron \cite{Rehacek:2004aa}.

While these approaches all require a combination of measurements 
with incompatible observables, it is also possible to obtain the complete
state information from a single measurement
performed on a larger Hilbert space, provided the state information
is first redistributed into this extended space \cite{DAriano:2002aa,Caves:2002aa, Du:2006aa, Allahverdyan:2004aa}. 
Ref. \cite{DAriano:2002aa} shows that it is possible to estimate the expectation
values of all observables of a quantum system by measuring only a single
``universal" observable on an extended Hilbert space.
In the Hilbert space of the quantum system, the reduced operator of this  
``universal observable" constitutes a minimal informationally complete 
positive operator-valued measurement \cite{Caves:2002aa}. 
Du et al. \cite{Du:2006aa} demonstrated an experimental example on this. 
Allahverdyan et al. \cite{Allahverdyan:2004aa} determine the conditions for
making this type of measurement robust by maximizing the determinant 
of the mapping between the quantum state and the measurement results. 

The possibility of obtaining the full quantum state information from a 
single measurement appears highly attractive and may well have
practical advantages since it avoids some experimental uncertainties
related to measurement setups for incompatible observables.
It requires, however, the redistribution of the information within the
extended Hilbert space.
This is achieved by coupling the system $S$, whose state is to be determined,
to an assistant $A$ and letting the combined system evolve for a 
suitable period. The sketch of this measurement idea is shown 
in Fig. \ref{SingleMeas} b).
As we show below, the success of the resulting measurements 
depends on the form of the Hamiltonian as well as on the duration
of the evolution and the choice of the final measurement on the 
combined system.

In this paper, we study the details of this type of measurements
using a (nuclear) spin 1/2 as the system whose quantum state
is to be determined, and a different spin 1/2 as the assistant.
We consider in detail what types of Hamiltonian can be used
to couple the system to the assistant, how the information
content of the resulting state can be maximized, and under what
conditions the scheme will fail.
As an experimental example, we present results from a nuclear
spin system, using nuclear magnetic resonance (NMR).

\section{Coupling system and assistant}

\subsection{Hamiltonian}

We consider two qubits interacting with local magnetic fields and
coupled through the Heisenberg interaction. The system Hamiltonian
can then be written as
\begin{equation}
\begin{array}{lll}
\hat{H} & = & \hat{H}_{z}(B_{1},B_{2})+\hat{H}_{ex}(J_{x},J_{y},J_{z}) \\
 & = & B_{1}\hat{S}_{z}^{1}+B_{2}\hat{S}_{z}^{2}+\\
& &
J_{x}\hat{S}_{x}^{1}\hat{S}_{x}^{2}+J_{y}\hat{S}_{y}^{1}\hat{S}_{y}^{2}+J_{z}\hat{S}_{z}^{1}\hat{S}_{z}^{2},
\end{array}
\label{e.H}
\end{equation}
where $\hat{S}_{\nu}^{k}=\frac{1}{2}\sigma_{\nu}^{k} (\nu=x,y,z)$
denotes the local spin operator for qubit $k$. 
The $B_{k}$s are the
strengths of the external magnetic fields (along the z axis) acting on
qubit $k$, and the $J_{\nu}$s are the Heisenberg exchange
constants.

For arbitrary $J_{\nu}$, this is often called the anisotropic
Heisenberg XYZ model. 
Some special cases are:
\begin{itemize}
\item {XXX (or isotropic Heisenberg): $J_{x}=J_{y}=J_{z}$}
\item{XXZ : $J_{x}=J_{y}\neq J_{z}$}
\item{XY : $J_{z}=0$}
\item{XZ : $J_{y}=0$}
\item{Heisenberg-Ising : $J_{x}=J_{y}=0$}
\end{itemize}

$J_{\nu}>0$ and $J_{\nu}<0$
correspond to the antiferromagnetic and ferromagnetic cases,
respectively. In many solid-state systems, the coupling constants
$J_{\nu}$ can be tuned by external fields and many proposals for
solid-state quantum information processors rely on their
tunability.

The Hamiltonian of Eq (\ref{e.H}) splits into three mutually commuting parts,
\begin{equation}
\hat{H}=\hat{H}_{zz}+\hat{H}_{0}+\hat{H}_{2}
\end{equation}
where
\begin{equation}
\begin{array}{l}
\hat{H}_{zz} = J_{z}\hat{S}_{z}^{1}\hat{S}_{z}^{2} \\
\hat{H}_{0} = B\gamma_B(\hat{S}_{z}^{1}-\hat{S}_{z}^{2})
+\frac{J}{2}(\hat{S}_{+}^{1}\hat{S}_{-}^{2}+\hat{S}_{-}^{1}\hat{S}_{+}^{2}) \\
\hat{H}_{2} =B(\hat{S}_{z}^{1}+\hat{S}_{z}^{2})
+\frac{J}{2} \gamma_J(\hat{S}_{+}^{1}\hat{S}_{+}^{2}+\hat{S}_{-}^{1}\hat{S}_{-}^{2}).
\end{array}
\end{equation}
$B=(B_{1}+B_{2})/2$ and $J=(J_{x}+J_{y})/2$ are the average field
and the coupling constant, and $\gamma_B=(B_{1}-B_{2})/(B_{1}+B_{2})$
and $\gamma_J=(J_{x}-J_{y})/(J_{x}+J_{y})$ are anisotropy
parameters. $\hat{S}_{\pm}^{k}=\hat{S}_{x}^{k}+i\hat{S}_{y}^{k}$
are the raising and lowering operators.

With this decomposition, the eigenvalues and eigenvectors
can be easily calculated by diagonalizing the subspaces
consisting of $\hat{H}_{0}$ and $\hat{H}_{2}$.
We obtain for the eigenvalues
\begin{equation}
\begin{array}{l}
\lambda_{1}=\frac{1}{4}J_{z}+\eta_{1},\\
\lambda_{2}=-\frac{1}{4}J_{z}+\eta_{2},\\
\lambda_{3}=-\frac{1}{4}J_{z}-\eta_{2},\\
\lambda_{4}=\frac{1}{4}J_{z}-\eta_{1},
\end{array}
\label{Eigval}
\end{equation}
and for the eigenvectors
\begin{equation}
\begin{array}{cc}
\left\vert \psi_{1}\right\rangle =\left(\begin{array}{c}
\cos(\theta_{1}/2)\\
0\\
0\\
\sin(\theta_{1}/2)\end{array}\right) & \left\vert
\psi_{2}\right\rangle =\left(\begin{array}{c}
0\\
\cos\frac{\theta_{2}}{2}\\
\sin\frac{\theta_{2}}{2}\\
0\end{array}\right)\\
\left\vert \psi_{3}\right\rangle =\left(\begin{array}{c}
0\\
-\sin\frac{\theta_{2}}{2}\\
\cos\frac{\theta_{2}}{2}\\
0\end{array}\right) & \left\vert \psi_{4}\right\rangle =\left(\begin{array}{c}
-\sin(\theta_{1}/2)\\
0\\
0\\
\cos(\theta_{1}/2)\end{array}\right).
\end{array}\label{EigVec}\end{equation}
Here
\begin{equation}
\begin{array}{l}
\eta_{1}=\sqrt{B^{2}+(J\gamma_J/2)^{2}} \\
\eta_{2}=\sqrt{(B\gamma_B)^{2}+(J/2)^{2}}
\end{array}.
\end{equation}
and
\begin{equation}
\begin{array}{l}
\cos\frac{\theta_{1}}{2}=\sqrt{\frac{\eta_{1}+B}{2\eta_{1}}}\\
\sin\frac{\theta_{1}}{2}=\frac{J\gamma_J/2}{\sqrt{2\eta_{1}(\eta_{1}+B)}}=sgn(J\gamma_J)\sqrt{\frac{\eta_{1}-B}{2\eta_{1}}}\\
\cos\frac{\theta_{2}}{2}=\sqrt{\frac{\eta_{2}+B\gamma_B}{2\eta_{2}}}\\
\sin\frac{\theta_{2}}{2}=\frac{J/2}{\sqrt{2\eta_{2}(\eta_{2}+B\gamma_B)}}=sgn(J)\sqrt{\frac{\eta_{2}-B\gamma_B}{2\eta_{2}}}
\end{array}.
\label{Paras}
\end{equation}

\subsection{Evolution}

We write the evolution operator as a product of the evolutions
generated by the three mutually commuting terms of Eq. (2):
\begin{equation}
\hat{U}(\tau)=e^{-i\hat{H}\tau}
=\hat{U}_{zz}(\tau)\hat{U}_{0}(\tau)\hat{U}_{2}(\tau)
\label{e.U}
\end{equation}
where
\begin{equation}
\begin{array}{lll}
\hat{U}_{zz}(\tau) & = & e^{-i\hat{H}_{zz}\tau}\\
                   & = & \cos(\frac{J_z\tau}{4})\mathbf{1}-i \sin(\frac{J_z\tau}{4})(4\hat{S}^{1}_z\hat{S}^{2}_z)\\
\hat{U}_{0}(\tau) & = & e^{-i\hat{H}_0\tau}\\
                   & = & \frac{1+\cos(\eta_2 \tau)}{2}\mathbf{1}+\frac{1-\cos(\eta_2 \tau)}{2}(4\hat{S}^{1}_z\hat{S}^{2}_z)+\\
                   &   & i2\cos\theta_2\sin(\eta_2\tau)(\hat{S}^{1}_z-\hat{S}^{2}_z)+\\
                   &   & i\sin\theta_2\sin(\eta_2\tau)(\hat{S}_{+}^{1}\hat{S}_{-}^{2}+\hat{S}_{-}^{1}\hat{S}_{+}^{2})\\
\hat{U}_{2}(\tau) & = & e^{-i\hat{H}_2\tau}\\
                  & = & \frac{1+\cos(\eta_1 \tau)}{2}\mathbf{1}-\frac{1-\cos(\eta_1 \tau)}{2}(4\hat{S}^{1}_z\hat{S}^{2}_z)+\\
                   &   & i2\cos\theta_1\sin(\eta_1\tau)(\hat{S}^{1}_z+\hat{S}^{2}_z)+\\
                   &   & i\sin\theta_1\sin(\eta_1\tau)(\hat{S}_{+}^{1}\hat{S}_{+}^{2}+\hat{S}_{-}^{1}\hat{S}_{-}^{2}) . \\
\end{array}
\label{U_comp}
\end{equation}

In the following, we will use a different operator basis for the diagonal terms:
we define the polarization operators
$I_{i}^{\alpha,\beta}=\frac{1}{2}\mathbf{1}\pm\hat{S}_{z}^{(i)}$.
In terms of these operators, the total propagator becomes
\begin{eqnarray}
\hat{U}(\tau)=a_{1}I_{1}^{\alpha}I_{2}^{\alpha}+a_{2}I_{1}^{\alpha}I_{2}^{\beta}+a_{3}I_{1}^{\beta}I_{2}^{\alpha}+a_{4}I_{1}^{\beta}I_{2}^{\beta} \nonumber\\
+d(\hat{S}_{+}^{1}\hat{S}_{-}^{2}+\hat{S}_{-}^{1}\hat{S}_{+}^{2})+b(\hat{S}_{+}^{1}\hat{S}_{+}^{2}+\hat{S}_{-}^{1}\hat{S}_{-}^{2}) ,
\label{U_para}
\end{eqnarray}
where
\begin{equation}
\begin{array}{l}
a_{1}=\cos^{2}\frac{\theta_{1}}{2}e^{-i\lambda_{1}\tau}+\sin^{2}\frac{\theta_{1}}{2}e^{-i\lambda_{4}\tau}\\
a_{2}=\cos^{2}\frac{\theta_{2}}{2}e^{-i\lambda_{2}\tau}+\sin^{2}\frac{\theta_{2}}{2}e^{-i\lambda_{3}\tau}\\
a_{3}=\sin^{2}\frac{\theta_{2}}{2}e^{-i\lambda_{2}\tau}+\cos^{2}\frac{\theta_{2}}{2}e^{-i\lambda_{3}\tau}\\
a_{4}=\sin^{2}\frac{\theta_{1}}{2}e^{-i\lambda_{1}\tau}+\cos^{2}\frac{\theta_{1}}{2}e^{-i\lambda_{4}\tau}\\
b=\frac{1}{2}\sin\theta_{1}(e^{-i\lambda_{1}\tau}-e^{-i\lambda_{4}\tau})\\
d=\frac{1}{2}\sin\theta_{2}(e^{-i\lambda_{2}\tau}-e^{-i\lambda_{3}\tau}) .
\end{array}
\label{Uparams}
\end{equation}

As we show in the following section, the evolution of Eq. (\ref{e.U})
transfers information between qubits in such a way that it becomes possible
to measure the complete quantum state of one qubit with a single apparatus,
as proposed by Allahverdyan et al.\cite{Allahverdyan:2004aa}.

\section{Measurement procedure}

\subsection{Principle}

Consider a two-level system \textbf{\emph{S}} (spin-$\frac{1}{2}$)
whose state can be represented by $\hat{\rho}=\left(
\begin{array}{cc}
\rho_{11} & \rho_{12} \\
\rho_{21} & \rho_{22} \\
\end{array}
\right)$ with the normalization $\rho_{11}+\rho_{22}=1$. To
determine the state $\hat{\rho}$, we can measure the vector
$\vec{s}= 2 Tr(\vec{\hat{S}}\hat{\rho})=(s_{x}, s_{y}, s_{z})^{T}$
where $s_x=\rho_{12}+\rho_{21}$, $s_y=i(\rho_{12}-\rho_{21})$,
$s_z=\rho_{11}-\rho_{22}$ and
$\vec{\hat{S}}=(\hat{S}_x,\hat{S}_y,\hat{S}_z)^T$. 

To transfer part of the state information to the assistant $A$, we couple
the two subsystems with the interaction Hamiltonian
$\hat{H}$ of Eq.(\ref{e.H}). 
At the time $t=0$,
the composite system \textbf{\emph{S+A}} is in the state
$\hat{\varrho}_{0}=\hat{\rho}^{(S)}\otimes \hat{\xi}^{(A)}$.
Without loss of generality, we assume
$\hat{\xi}=\frac{1}{2}\mathbf{1}+\epsilon S_z, (0 \leq\epsilon\leq
1)$. Here, the superscripts $S$ and $A$ refer to the two subsystems.

Under the effect of the coupling
Hamiltonian of Eq. (1), this state evolves into
$\hat{\varrho}_{\tau} = \hat{U}(\tau) \, \hat{\varrho}_{0} \, \hat{U}^{\dag}(\tau)$.
On this state, we perform repeatedly
measure the simplest possible nondegenerate, factorized observable 
$\hat{\Omega} =\sum_{\alpha=1} ^4 \varpi_\alpha \hat{P}_\alpha $, 
which determines the
complete set $\{P_\alpha \}$ of the joint probabilities, 
They correspond to the eigenvalues of $\hat{\varrho}_{\tau}$ in the
eigenbasis of the observable $\hat{\Omega} $. Since these values were generated
from the initial state 
by a one-to-one mapping, $P_\alpha=P_{kq}=\sum_{ij} \mathcal{M}_{kq,ij}
\rho_{ij}$, we can invert this mapping to calculate the original
state $\hat{\rho}$ \cite{Allahverdyan:2004aa}.

The precision of the back-calculation depends on the size of the
determinant $\Delta=\det(M_{kq,ij})$: If $|\Delta|$ is small, any
(experimental) error in the measurement of  $P_{kq}$ will
result in a large error in $\rho_{ij}$, roughly $\propto
1/|\Delta|$. We therefore seek to maximize $|\Delta|$ and thereby
the precision of the measurement. This maximization is achieved by
a suitable choice of the Hamiltonian $\hat{H}$, the duration
$\tau$, and the single observable $\hat{\Omega} $.

\subsection{Symmetry properties of the evolution}

\begin{figure}[tbh]
\begin{centering}
\includegraphics[width=0.6\columnwidth]{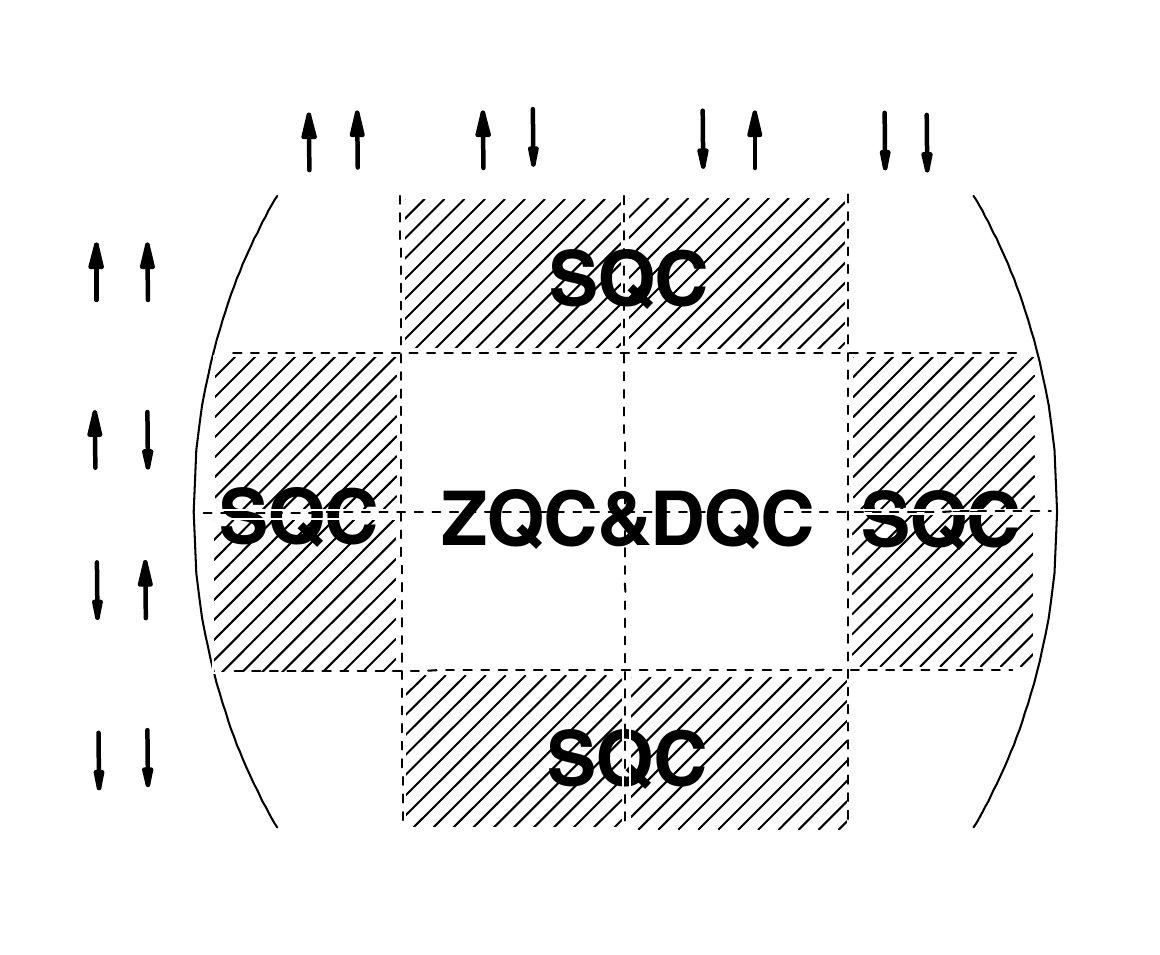}
\end{centering}
\caption{Subspaces of the density operator that are invariant
by the evolution under the Hamiltonian of Eq. (\ref{e.H}).}
\label{space}
\end{figure}

The Hamiltonian $\hat{H}$ of Eq. (\ref{e.H}) consists of
three commuting parts:  $\hat{H}_{zz}$, $\hat{H}_{0}$,
and $\hat{H}_{2}$.
All of these terms are invariant under $\pi$-rotations around
the z-axis.
We use this property to separate the density operator into two parts
that transform irreducibly under this symmetry operation:
One part which is invariant with respect to $\pi_z$ rotations,
and the second part, which changes sign.
The first part includes diagonal terms, zero quantum and double quantum coherence
\cite{ErnstBook:1994aa};
the second part includes the single quantum coherence terms.
The symmetry of the Hamiltonian implies that the evolution does not transfer
information from one subspace to the other.
Figure \ref{space} illustrates the division of the density operator into these subspaces.

A consequence of this separation of the system into two distinct
subspaces is that it restricts the possible choice of observables.
In particular, if we choose the z-components of the two spins to
the single observable $\hat{\Omega} $, then all possible
combinations fall into the subspace that is invariant under
$\pi_z$ rotations and therefore does not provide information about
the other subspace.
For this paper, we choose the x-components of both spins;
another, equivalent choice would be the y-components.

\subsection{Transfer matrix}

This evolution process, together with the subsequent measurement of 
the single observable $\hat{\Omega} $,
transfers the information from the initial state
$\hat{\varrho}_{0}$ to the set of measurement results $P_{\alpha}=P_{kq}$,
which are the expectation values of the operators
$$
\hat{P}_{kq} = (\frac{1}{2}\mathbf{1}- (-1)^{k}\hat{S}_{x}^{(S)})
\otimes (\frac{1}{2}\mathbf{1}- (-1)^{q}\hat{S}_{x}^{(A)})
$$
acting on the state
$\hat{\varrho}_{\tau}$:
$$
P_{kq} = Tr \{ \hat{P}_{kq} \hat{U}(\tau) \hat{\varrho}_{0}
\hat{U}^{\dagger} (\tau) \} .
$$
The indices $k,q$ can have the values 1, 2.

We use the transfer matrix $\mathcal{M}$ to describe
this map:
\begin{equation}
\mathcal{M} \left(\begin{array}{c}
\rho_{11}\\
\rho_{12}\\
\rho_{21}\\
\rho_{22}\end{array}\right)=\left(\begin{array}{c}
P_{11}\\
P_{12}\\
P_{21}\\
P_{22}
\end{array}\right) .
\label{mapM}
\end{equation}

Its elements are
\begin{equation}
\begin{array}{l}
\mathcal{M}_{11,11}=\mathcal{M}_{22,11}=\frac{1}{8}((1+ \epsilon )|a_{1}+b|^2+(1- \epsilon )|a_{2}+d|^2)\\
\mathcal{M}_{12,11}=\mathcal{M}_{21,11}=\frac{1}{8}((1+ \epsilon )|a_{1}-b|^2+(1- \epsilon )|a_{2}-d|^2)\\
\mathcal{M}_{11,12}=-\mathcal{M}_{22,12}=\mathcal{M}_{11,21}^{*}=-\mathcal{M}_{22,21}^{*}\\
=\frac{1}{8}((1+ \epsilon )(a_{1}+b)(a_{3}^{*}+d^{*})+(1- \epsilon )(a_{2}+d)(a_{4}^{*}+b^{*})\\
\mathcal{M}_{22,22}=-\mathcal{M}_{21,12}=\mathcal{M}_{12,21}^{*}=-\mathcal{M}_{33}^{*}\\
=\frac{1}{8}((1+ \epsilon )(a_{1}-b)(a_{3}^{*}-d^{*})+(1- \epsilon )(a_{2}-d)(a_{4}^{*}-b^{*})\\
\mathcal{M}_{11,22}=\mathcal{M}_{22,22}=\frac{1}{8}((1+ \epsilon )|a_{3}+d|^2+(1- \epsilon )|a_{4}+b|^2)\\
\mathcal{M}_{12,22}=\mathcal{M}_{21,22}=\frac{1}{8}((1+ \epsilon )|a_{3}-d|^2+(1- \epsilon )|a_{4}-b|^2)\\
\end{array}
\label{e.elem.M}
\end{equation}
and its determinant is
\begin{equation}
\Delta =8
\Im(\mathcal{M}_{12,12}^{*}\mathcal{M}_{11,12})(\mathcal{M}_{11,11}\mathcal{M}_{12,22}-\mathcal{M}_{12,11}\mathcal{M}_{11,22}) .
\label{e.detM}
\end{equation}
Here $\Im(c)$ denotes the imaginary part of $c$. Using Eqs.
(\ref{Eigval}), (\ref{Paras}), (\ref{Uparams}) and (\ref{e.elem.M}),
we find for its absolute value
\begin{equation} \label{e.Delta}
\begin{array}{ll}
|\Delta|=&\frac{1}{32}|(1-\epsilon^2)\sin(-J_{z}\tau)\{[\sin(2\theta_{1})\sin^{2}(\eta_{1}\tau)]^{2}-\\
&[\sin(2\theta_{2})\sin^{2}(\eta_{2}\tau)]^{2}\}+2\epsilon [\sin(2\theta_{1})\sin^{2}(\eta_{1}\tau)\\
&+\sin(2\theta_{2})\sin^{2}(\eta_{2}\tau)]\{[1-2\sin^{2}\theta_{1}\sin^{2}(\eta_{1}\tau)]\\
&\times \sin\theta_{2}\sin(2\eta_{2}\tau)-[1-2\sin^{2}\theta_{2}\sin^{2}(\eta_{2}\tau)]\\
&\times \sin\theta_{1}\sin(2\eta_{1}\tau)\}| .
\end{array} 
\end{equation}

\section{Optimization: Maximizing $|\Delta|$}

The size of the determinant  $|\Delta|$ of the transfer mapping
determines the quality of measurement. Maximizing $|\Delta|$ will
minimize the statistical error of the estimation during the
inversion of Eq. (\ref{mapM}). We can maximize it by an
appropriate choice of the initial condition of the assistant, the
parameters of the Hamiltonian that generates the evolution $U$,
and the duration of the evolution.

\begin{figure}[tbh]
\begin{centering}
\includegraphics[width=0.8\columnwidth]{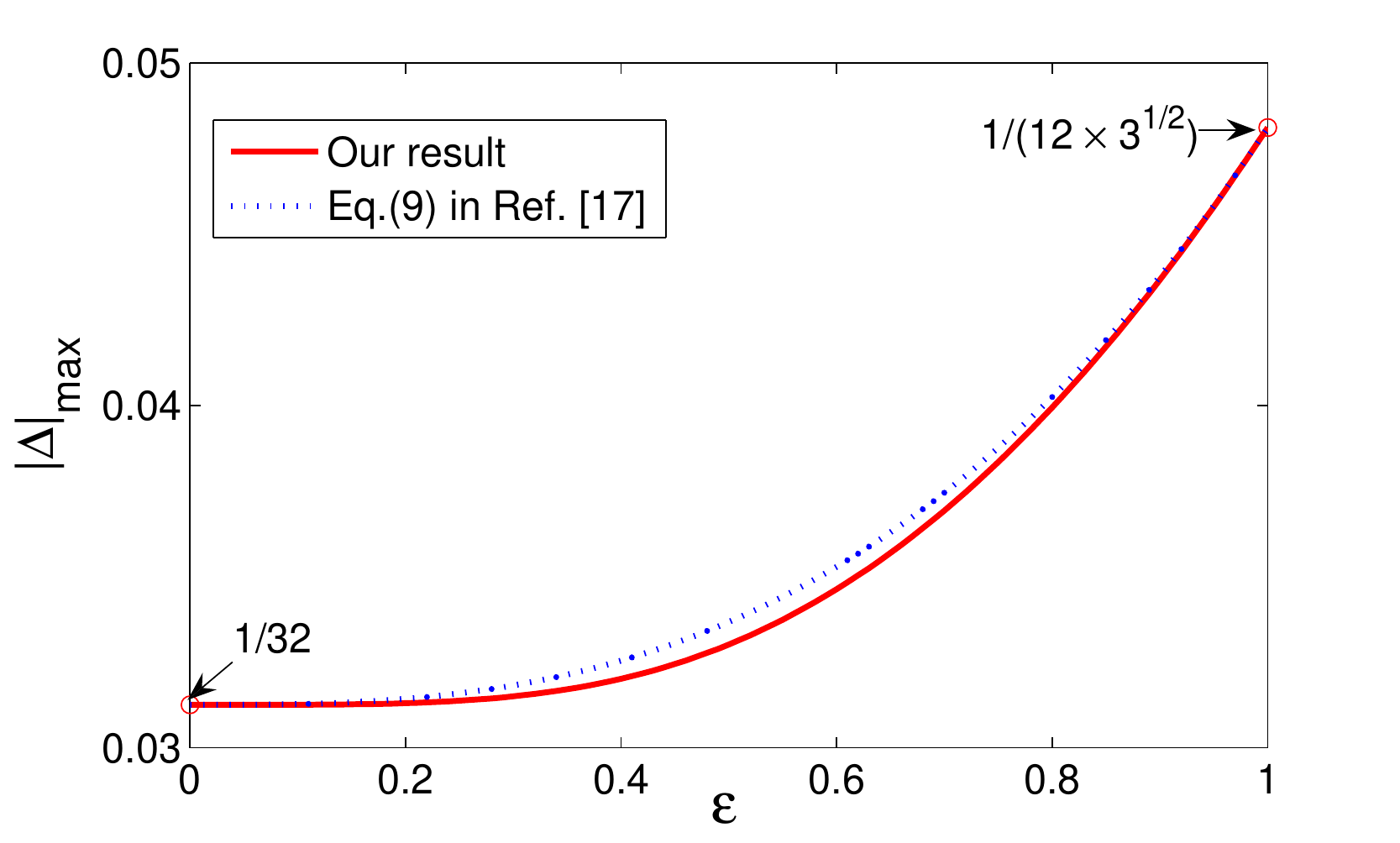}
\end{centering}
\caption{(Color Online) Maximal determinant size
$|\Delta|_{max}$ versus the polarization $\epsilon$ by the
Heisenberg exchange interaction (\ref{e.H}), compared to the 
general case of arbitrary exchange interaction 
(Eq. (9)) in Ref. \cite{Allahverdyan:2004aa}).}
\label{Dmax}
\end{figure}

Figure \ref{Dmax} plots the maximum possible determinant size
$|\Delta|_{max}$ for the Hamiltonian of Eq. (\ref{e.H}) as a
function of the polarization $\epsilon$ of the assistant. 
Clearly, the quality of
the measurement should increase with increasing polarization of
the assistant. The dashed line in Fig. \ref{Dmax} shows for
comparison the maximum possible value for a general exchange
interaction, taken from Ref. \cite{Allahverdyan:2004aa}. At the extreme cases
of zero and full polarization, the Heisenberg coupling Hamiltonian
allows one to reach the maximum possible value, but for
intermediate polarizations, its maximum value is slightly lower
than for the general case.

Let us now focus on two extreme situations $\epsilon=1$ (\emph{a
pure state}) and $\epsilon=0$ (\emph{a completely disordered
state}).

\subsection{Assistant in pure state}

When the assistant \emph{\textbf{A}} starts in \emph{a pure state}
$\hat{\xi}^{(A)} = \frac{1}{2}\mathbf{1}+S_z^{(A)}$ (corresponding to $\epsilon=1$),
the determinant becomes
\begin{equation}
\begin{array}{ll}
|\Delta|=&\frac{1}{16}|
[\sin(2\theta_{1})\sin^{2}(\eta_{1}\tau)+\sin(2\theta_{2})\sin^{2}(\eta_{2}\tau)]\\
&\times \{[1-2\sin^{2}\theta_{1}\sin^{2}(\eta_{1}\tau)]\sin\theta_{2}\sin(2\eta_{2}\tau)\\
&-[1-2\sin^{2}\theta_{2}\sin^{2}(\eta_{2}\tau)]\sin\theta_{1}\sin(2\eta_{1}\tau)\}|.
\end{array}
\end{equation}
We can see from this expression that $|\Delta|$ is independent of
the coupling strength $J_z$ along the z-axis. A Heisenberg XY
interaction is therefore sufficient for optimizing the evolution.
We therefore specialize to this case. Using the substitutions
\begin{equation}\label{subs}
\begin{array}{l}
  \sin\frac{\Xi_k}{2}=\sin\theta_k \sin(\eta_k\tau)  \\
  \sin\Lambda_k=\cos(\eta_k\tau)/\cos\frac{\Xi_k}{2},(k=1,2),
\end{array}
\end{equation}
we rewrite $|\Delta|$ as
\begin{equation}
\begin{array}{ll}
|\Delta|=&\frac{1}{16}|(\sin\Xi_{1}\cos\Lambda_1+\sin\Xi_{2}\cos\Lambda_2) \\
&\times
(\cos\Xi_{1}\sin\Xi_2\sin\Lambda_2-\cos\Xi_2\sin\Xi_{1}\sin\Lambda_1)|
.
\end{array}
\end{equation}
In terms of these parameters, an optimal solution ($|\Delta| = 1/(12\sqrt{3})$)
is given by the following set of parameters:
\begin{equation}
\begin{array}{l}
\Lambda_1=\Lambda_2=\Lambda \\
\sin(2\Lambda)= \pm 1\\
\sin\frac{\Xi_1-\Xi_2}{2}= \pm \frac{1}{\sqrt{3}} \\
\sin\frac{\Xi_1+\Xi_2}{2}= \pm 1 .
\end{array} 
\end{equation}
This parameter set corresponds to the following parameters
of the Hamiltonian (1):
\begin{equation}
\label{Con_pure}
\begin{array}{l}
\eta_{k}\tau = m\pi \pm \frac{1}{2} \arccos ( -\Gamma_k)\\
B=\pm \eta_1\sqrt{(1-\Gamma_1)/(1+\Gamma_1)}\\
\gamma_B=\pm
\sqrt{\frac{(1+\Gamma_1)(1-\Gamma_2)}{(1-\Gamma_1)(1+\Gamma_2)}}\frac{\eta_2}{\eta_1} \\
J=\pm 4\eta_2\sqrt{2\Gamma_2/(1+\Gamma_2)} \\
\gamma_J=\pm
\sqrt{\frac{\Gamma_1(1+\Gamma_2)}{\Gamma_2(1+\Gamma_1)}}\frac{\eta_1}{\eta_2} .
\end{array}
\end{equation}
Here, $m$ is an integer and $\Gamma_k=\sin^2
\frac{\Xi_k}{2},(k=1,2)$. $(\Gamma_1,\Gamma_2)$ take the pairs of
values
$(\frac{1}{2}-\frac{\sqrt{3}}{6},\frac{1}{2}+\frac{\sqrt{3}}{6})$
or
$(\frac{1}{2}+\frac{\sqrt{3}}{6},\frac{1}{2}-\frac{\sqrt{3}}{6})$.
Without loss of generality, we set $\tau=1$. 
The optimal Hamiltonian is then
\begin{equation}
H^{opt}=1.1458\hat{S}_{z}^{1}-0.2935\hat{S}_{z}^{2}+3.3820\hat{S}_{x}^{1}
\hat{S}_{x}^{2}-1.2747\hat{S}_{y}^{1}\hat{S}_{y}^{2}.
\end{equation}

\subsection{Completely disordered assistant.}

When the assistant \emph{\textbf{A}} is initially in \emph{a completely
disordered state} $\hat{\xi}^{(A)}=\frac{1}{2}\mathbf{1}$ ($\epsilon=0$),
we have
\begin{equation}
\begin{array}{ll}
|\Delta|=&\frac{1}{32}|\sin(-J_{z}\tau)||\{[(\sin(2\theta_{1})\sin^{2}(\eta_{1}\tau)]^{2}\\
&-[(\sin(2\theta_{2})\sin^{2}(\eta_{2}\tau)]^{2}\}|.
\end{array}
\end{equation}
$|\Delta|$ reaches its maximum of $1/32$ when
\begin{equation}
 \sin(J_{z}\tau)=\pm
1\Rightarrow J_{z}\tau=\frac{\pi}{2}(2n-1)
\end{equation}
($n$ integer) and simultaneously
\begin{equation}
\sin(2\theta_{1})\sin^{2}(\eta_{1}\tau) = \pm 1
\mbox{  and  }
\sin(2\theta_{2})\sin^{2}(\eta_{2}\tau) = 0
\label{e23}
\end{equation}
or
\begin{equation}
\sin(2\theta_{1})\sin^{2}(\eta_{1}\tau) = 0
\mbox{  and  }
\sin(2\theta_{2})\sin^{2}(\eta_{2}\tau) = \pm 1 .
\label{e24}
\end{equation}

Condition (\ref{e23}) corresponds to the following parameters for the Hamiltonian:
\begin{equation}
 \begin{array}{ccl}
|B|\tau & = &|2m-1|  \frac{\sqrt{2}\pi}{4}\\
|J_x-J_y| & = & 4 |B|\\
\gamma_B  & = & 0 \mbox{ OR }J_x+J_y=0 \mbox{ OR }\eta_{2}\tau=l\pi  \\
\end{array}
\end{equation}
and (\ref{e24}) to
\begin{equation}
 \begin{array}{ccl}
|J|\tau & = & |2m-1| \frac{\sqrt{2}\pi}{2}\\
|B_1-B_2|  & = & |J|\\
B_1+B_2  & = & 0 \mbox{ OR }\gamma_J=0 \mbox{ OR }\eta_{1}\tau=l\pi ,
\end{array}
\end{equation}
where $m,l$ are integers.

These relationships define
six classes of Heisenberg exchange interactions 
that optimally transfer information
from the system to the combined system plus assistant.
The transfer is determined by the product of the Hamiltonian 
and the evolution time $\tau$.
Without loss of generality, we choose $\tau=\frac{\pi}{4}$.
In these units, some possibilities are
\begin{description}
\item [{(a)}] XYX model:
$H^{opt}=\sqrt{2}(\hat{S}_{z}^{1}+\hat{S}_{z}^{2})
+2(\hat{S}_{x}^{1}\hat{S}_{x}^{2}+\hat{S}_{z}^{1}\hat{S}_{z}^{2})
+2(1-2\sqrt{2})\hat{S}_{y}^{1}\hat{S}_{y}^{2}$, as shown in
Ref.\cite{Allahverdyan:2004aa};
\item [{(b)}] XXZ model:
$H^{opt}=\sqrt{2}(\hat{S}_{z}^{1}-\hat{S}_{z}^{2})
+2\sqrt{2}(\hat{S}_{x}^{1}\hat{S}_{x}^{2}+\hat{S}_{y}^{1}\hat{S}_{y}^{2})
+2\hat{S}_{z}^{1}\hat{S}_{z}^{2}$;
\item [{(c)}] XZ model:
$H^{opt}=\sqrt{2}(\hat{S}_{z}^{1}\pm \hat{S}_{z}^{2})
+4\sqrt{2}\hat{S}_{x}^{1}\hat{S}_{x}^{2}
+2\hat{S}_{z}^{1}\hat{S}_{z}^{2}$.
\end{description}

\section{Failure analysis}

The measurement scheme fails when  $\Delta=0$.
From Eq. (\ref{e.Delta}), we see that this occurs when
\begin{equation}\label{Delta0}
\sin(2\theta_{1})\sin^{2}(\eta_{1}\tau)+\sin(2\theta_{2})\sin^{2}(\eta_{2}\tau)=0
\end{equation}
or when
\begin{equation}
\begin{array}{l}
(1-\epsilon^2)\sin(-J_{z}\tau)[\sin(2\theta_{1})\sin^{2}(\eta_{1}\tau)
-\sin(2\theta_{2})\sin^{2}(\eta_{2}\tau)]\\
+2\epsilon\{[1-2\sin^{2}\theta_{1}\sin^{2}(\eta_{1}\tau)]
\sin\theta_{2}\sin(2\eta_{2}\tau)\\
-[1-2\sin^{2}\theta_{2}\sin^{2}(\eta_{2}\tau)]
\sin\theta_{1}\sin(2\eta_{1}\tau)\}=0 ,
\end{array} ,
\end{equation}
independent of the initial state of the assistant \textbf{\emph{A}}.

A simple case is $\sin(2\theta_1)=\sin(2\theta_2)=0$,
i.e., $J=0$ or $B=0$ or ($\gamma_B=0$ and $\gamma_J=0$).
These cases correspond, e.g., to
\begin{itemize}
\item{
a weakly-coupled
liquid-state NMR Hamiltonian ($J_{x}=J_{y}=0$)}
\item{any Heisenberg
interaction without external field}
\item{an isotropic Heisenberg
interaction in the XY plane in a uniform external field.}
\end{itemize}

In all of these cases, the resulting evolution
cannot generate a state that allows one to measure the complete information.

Another case that fulfills Eq. (\ref{Delta0}) is
\begin{equation}
\sin(2\theta_{1})=-\sin(2\theta_{2})\Rightarrow\frac{\gamma_J}{\gamma_B}=-\frac{\eta_1^2}{\eta_2^2}
\end{equation}
and
\begin{equation}
|\sin(\eta_{1}\tau)|=|\sin(\eta_{2}\tau)|\Rightarrow\eta_{1}\tau=|m\pi\pm\eta_{2}\tau|.
\end{equation}
If, e.g., $\eta_1=\eta_2$, we get the condition
\begin{equation}\label{FailCon}
(\gamma_B,\gamma_J)=(1,\pm 1) \mbox{ or } (-1,1)
\end{equation}
for $\Delta=0$.

From Eq. (\ref{e.Delta}), we can seen that for $\epsilon=1$ (a
pure state), $|\Delta|$ does not depend on $J_z$, while for
$\epsilon=0$ (completely disordered state), $|\Delta|$
depends strongly on $J_z$.
For $\epsilon=0$, it is obvious that $\Delta=0$
when
\begin{equation}
\sin(J_{z}\tau)=0\Rightarrow J_{z}\tau=n\pi.
\end{equation}
Hence, the existence of the coupling along the z-axis (i.e.,$J_{z}\neq0$)
is a necessary condition for this measurement scheme when the assistant
\textbf{\emph{A}} is initially prepared in a completely disordered
state. In this case, the failure condition (\ref{FailCon}) can be
further modified to
\begin{equation}
(\gamma_B,\gamma_J)=(\pm 1,\pm 1),
\end{equation}
which means that when any two among $J_{x},J_{y},B_{1},B_{2}$ are
equal to zero, the measurement scheme fails.

\section{Quantum Simulation of the Exchange Hamiltonian}

In liquid-state NMR systems, the natural Hamiltonian for a system of
two spins is
\begin{equation}
\hat{H}_{NMR}= \omega_{1}\hat{S}_{z}^{1} + \omega_{2}\hat{S}_{z}^{2}
+2\pi J_{12}\hat{S}_{z}^{1}\hat{S}_{z}^{2},
\label{e.H_nmr}
\end{equation}
where $\omega_{1,2}$ represent the Larmor angular frequencies of
the two qubits (in the rotating frame) and $J_{12}$ the spin-spin coupling constant.
This is equivalent to the Heisenberg-Ising model.
As discussed in the preceding section, this Hamiltonian cannot be used
to transfer the information, since the transfer matrix becomes singular,
$\Delta\equiv0$.
Therefore,
the key to implement this measurement scheme in a liquid-state NMR
system is to first perform a quantum simulation of the Hamiltonian
(\ref{e.H}).
We briefly discuss two techniques for realizing such an evolution.

\subsection{Short period expansion}

Assuming that we can realize parts of the Hamiltonian experimentally,
we write the total Hamiltonian as a sum,
$$
\hat{H}=\sum\limits _{k=1}^{L}\hat{H}_{k} .
$$
In general, the different terms do not commute with each other, and it
is therefore not sufficient to generate them sequentially.
However, if the evolution under each term is sufficiently short, it is possible
to approximate the overall evolution in this way.

Using a symmetrized version of the Trotter formula \cite{Trotter:1959aa},
$$
e^{(A+B)\tau} = e^{(A\tau/2)}e^{(B\tau)}e^{(A\tau/2)} + O(\tau^3)
$$
we expand the propagator as
\begin{eqnarray}
e^{-i\hat{H}\Delta t} \simeq [e^{-i\hat{H}_{1}\frac{\Delta t}{2}}e^{-i\hat{H}_{2}\frac{\Delta t}{2}}...e^{-i\hat{H}_{L}\frac{\Delta t}{2}}] \nonumber \\
\cdot  [e^{-i\hat{H}_{L}\frac{\Delta t}{2}} e^{-i\hat{H}_{L-1}\frac{\Delta t}{2}}...e^{-i\hat{H}_{1}\frac{\Delta t}{2}} ]+O(\Delta t^{3}) ,
\end{eqnarray}
which approximates the desired evolution to second order in $\Delta t$.
Keeping $\Delta t$ short enough, this allows one to efficiently simulate
the target Hamiltonian (\ref{e.H}) by concatenating these evolution periods
until the correct total evolution is reached.

Our target Hamiltonian can be decomposed into two non-commuting
parts $\hat{H}_z  +\hat{H}_{zz}$ and
$\hat{H}_{xy}=J_x\hat{S}^{1}_{x}\hat{S}^{2}_{x}+J_y\hat{S}^{1}_{y}\hat{S}^{2}_{y}$.
We thus generate the overall evolution (\ref{e.U}) as
\begin{equation}
\hat{U}(\tau)=\hat{U}^m(\Delta t)=[\hat{U}_{z}(\frac{\Delta
t}{2})\hat{U}_{xy}(\Delta t)\hat{U}_{z}(\frac{\Delta
t}{2})]^m+O(\Delta t^{3}),
\label{Imp_Udt}
\end{equation}
where $\tau=m\Delta t$ is the total duration, and
$$
\hat{U}_{z}(\frac{\Delta
t}{2})=e^{-i(\hat{H}_z+\hat{H}_{zz})\frac{\Delta t}{2}} ,
$$
and
$$
\hat{U}_{xy}(\Delta t)=e^{-i\hat{H}_{xy}\Delta t}
$$
represent the evolutions under the partial Hamiltonians.

Taking as an example the XZ model (case (c) in Sec. IV B), it is
sufficient to choose the number of evolution periods $m=2$ for $\tau
= \pi/4$: the resulting approximate evolution
$$
\hat{U}^{ap}(\tau)=[\hat{U}_{z}(\frac{\pi}{16})
\hat{U}_{xy}(\frac{\pi}{8})\hat{U}_{z}(\frac{\pi}{16})]^2
$$
with
$$
\hat{U}_{z}(\frac{\pi}{16})=e^{-i[(\sqrt{2}(\hat{S}^{1}_{z}\pm
\hat{S}^{2}_{z})+2\hat{S}^{1}_{z}\hat{S}^{2}_{z}]\frac{\pi}{16}}
$$
and
\begin{eqnarray}
\hat{U}_{xy}(\frac{\pi}{8}) & = & e^{-i\hat{S}^{1}_{x}
\hat{S}^{2}_{x}\frac{\sqrt{2}\pi}{2}} \nonumber \\
& = &  e^{-i(\hat{S}^{1}_{y}
+\hat{S}^{2}_{y})\frac{\pi}{2}}e^{-i\hat{S}^{1}_{z}
\hat{S}^{2}_{z}\frac{\sqrt{2}\pi}{2}}e^{i(\hat{S}^{1}_{y}
+\hat{S}^{2}_{y})\frac{\pi}{2}}
\label{Uxy}
\end{eqnarray}
has a fidelity of 0.9958 with the target evolution, where the
fidelity is defined as
$$
F(\hat{U}(\tau),\hat{U}^{ap}(\tau))=\frac{Tr(\hat{U}^{\dag}(\tau)\hat{U}^{ap}(\tau))}{4} .
$$
The $\hat{U}_z(\frac{\pi}{16})$ operator can be implemented by a free evolution
period  under the internal Hamiltonian if we choose
$\omega_1=\pm  \omega_2 = \sqrt{2} \pi J_{12}$
and set the duration to
$d_{1}=\frac{1}{16 \, J_{12}}$.

The $\hat{U}_{xy}(\frac{\pi}{8})$ operator can be implemented by four 
$\frac{\pi}{2}$ pulses  (corresponding to $e^{-i\hat{S}^{i}_{y}\frac{\pi}{2}} $) 
and a free precession period of duration $d_{2}=\frac{\sqrt{2}}{4 J_{12}}$.
This evolution period implements 
$e^{-i\hat{S}^{1}_{z}\hat{S}^{2}_{z}\frac{\sqrt{2}\pi}{2}}$;
we therefore refocus the chemical shift terms by inserting 
refocusing $\pi$ pulses in the middle of this period.
According to Eq. (\ref{Uxy}), the second set of $\pi/2$ pulses should
rotate the spins around the $-y$ axis.
Here, we choose the $+y$-axis instead to compensate for the inversion
of the axes system by the $\pi$-pulses.

The resulting pulse sequence that generates one segment of
$\hat{U}^{ap}(\tau)$ is
\begin{equation}
d_{1}-\left[\frac{\pi}{2}\right]_{y}^{1}\left[\frac{\pi}{2}\right]_{y}^{2}-\frac{d_{2}}{2}-\left[\pi\right]_{-y}^{1}\left[\pi\right]_{-y}^{2}-\frac{d_{2}}{2}-\left[\frac{\pi}{2}\right]_{y}^{1}\left[\frac{\pi}{2}\right]_{y}^{2}-d_{1},
\label{Unmrde}
\end{equation}
where $\left[\theta\right]_{\hat{\nu}}^{k}$ denotes a $\theta$
rotation of qubit $k$ around the $\hat{\nu}$ axis.

\subsection{Exact decomposition}

In some cases, it is possible to achieve the exact transformation by a suitable
decomposition of the evolution, using, e.g.,
$$
e^{-i\hat{R}\hat{H}\hat{R}^{\dag}\tau}=\hat{R}e^{-i\hat{H}\tau}\hat{R}^{\dag} .
$$
For the propagator (\ref{e.U}), we can use the decomposition
\begin{equation}
\hat{U}=\hat{R}\cdot e^{-i\hat{H}_{diag}\tau}\cdot\hat{R}^{\dagger} ,
\label{e.Udecom}
\end{equation}
where $\hat{H}_{diag}$ is the diagonal form of the Hamiltonian.
The transformation
\begin{equation}
\hat{R}=\left(
\begin{array}{llll}
\cos\frac{\theta_{1}}{2} &  &  & -\sin\frac{\theta_{1}}{2}\\
 & \cos\frac{\theta_{2}}{2} & -\sin\frac{\theta_{2}}{2}\\
 & \sin\frac{\theta_{2}}{2} & \cos\frac{\theta_{2}}{2}\\
\sin\frac{\theta_{1}}{2} &  &  & \cos\frac{\theta_{1}}{2}
\end{array}
 \right) ,
\end{equation}
which diagonalizes the Hamiltonian (\ref{e.H}),
can be implemented experimentally (up to an irrelevant overall phase factor)
by the pulse sequence
\begin{eqnarray}
\left[\frac{\pi}{2}\right]_{-y}^{1}\left[\frac{\pi}{2}\right]_{\varphi}^{2}-
\frac{\tau_1}{2}-\left[\pi\right]_{y}^{1}\left[\pi\right]_{-x}^{2}-
\frac{\tau_1}{2}-\left[\frac{\pi}{2}\right]_{-x}^{1}\left[\frac{\pi}{2}\right]_{y}^{2}  \nonumber\\
-\frac{\tau_2}{2}-\left[\pi\right]_{x}^{1}\left[\pi\right]_{-y}^{2}-
\frac{\tau_2}{2}-\left[\frac{\pi}{2}\right]_{-x}^{1}\left[\frac{\pi}{2}\right]_{-y}^{1}
\left[\frac{\pi}{2}\right]_{y}^{2}
\left[\frac{\pi}{2}\right]_{\varphi}^{2} \nonumber \\
\label{e.pulseR}
\end{eqnarray}
with $\tau_1=\frac{2|\theta_1-\theta_2|}{\pi J_{12}}$,
$\tau_2=\frac{2|\theta_1+\theta_2|}{\pi J_{12}}$ and $\varphi = x$
or $-x$ for $\theta_1>\theta_2$ or $\theta_1<\theta_2$, and
$\hat{R}^{\dagger}$ by the Hermite time-reversed sequence.

The evolution under the diagonal Hamiltonian
\begin{equation}
\begin{array}{l}
\hat{U}_{diag}=e^{-i\hat{H}_{diag}\tau}=\left(\begin{array}{llll}
e^{-i\lambda_{1}\tau}\\
 & e^{-i\lambda_{2}\tau}\\
 &  & e^{-i\lambda_{3}\tau}\\
 &  &  & e^{-i\lambda_{4}\tau}\end{array}\right)\\
 \end{array}
\end{equation}
is realized by the pulse sequence
\begin{equation}
\frac{\tau_{3}}{2}-\left[\pi\right]_{x}^{1}\left[\pi\right]_{x}^{2}
-\frac{\tau_{3}}{2}-\left[\frac{\pi}{2}\right]_{-x}^{1}\left[\frac{\pi}{2}
\right]_{-x}^{2}\left[\beta_{1}\right]_{y}^{1}\left[\beta_{2}
\right]_{y}^{2}\left[\frac{\pi}{2}\right]_{-x}^{1}
\left[\frac{\pi}{2}\right]_{-x}^{2} ,
\label{e.pulseUd}
\end{equation}
where
$\tau_{3}=\frac{\lambda_{1}-\lambda_{2}-\lambda_{3}+\lambda_{4}}{2\pi
J_{12}}\tau$,
$\beta_{1}=\frac{\lambda_{1}+\lambda_{2}-\lambda_{3}-\lambda_{4}}{2}\tau$
and
$\beta_{2}=\frac{\lambda_{1}-\lambda_{2}+\lambda_{3}-\lambda_{4}}{2}\tau$.
The first part of this sequence implements an evolution under the
J-coupling alone, the second part implements a composite
$z-$rotation of the two qubits by angles $\beta_1$ and $\beta_2$.

An alternative realization of $\hat{U}_{diag}$ is achieved by letting the system evolve
under a constant Hamiltonian with
$$
\omega_1 = \frac{\lambda_{1}+\lambda_{2}-\lambda_{3}
-\lambda_{4}}{2}\cdot\frac{\tau}{\tau_{3}}
$$
and
$$
\omega_2 = \frac{\lambda_{1}-\lambda_{2}+\lambda_{3}
-\lambda_{4}}{2}\cdot\frac{\tau}{\tau_{3}}
$$
for a period $\tau_{3}$.

For the example (c) in Sec. IV B, we choose the parameters
\begin{eqnarray}
\tau_1 & = & \tau_3  =\frac{1}{4 J_{12}},\tau_2  =  \frac{3}{4 J_{12}}, \nonumber\\
\beta_1& = & \frac{\pi (2+\sqrt{2})}{4}, \beta_2  =  \frac{\pi
(2-\sqrt{2})}{4}.
\end{eqnarray}
When $B_1=B_2$, $\varphi = x$ in the sequence (\ref{e.pulseR}); when
$B_1=-B_2$, $\varphi = -x$.

\section{EXPERIMENTAL IMPLEMENTATION}

Experiments were performed at room temperature on a Bruker Avance II 500
MHz spectrometer equipped with a Triple-Broadband-Observe(TBO) probe
at the frequencies 500.23 MHz for $^{1}$H and 125.13 MHz for $^{13}$C.
For the qubit system, we chose $^{13}$C-labelled chloroform diluted in acetone-d$_{6}$.
The ``unknown" state $\hat{\rho}$ was prepared on the spin of the proton
nuclei ($^{1}$H), which served as the quantum system \textbf{\emph{S}} (qubit 1),
and the spin of the $^{13}$C nuclei was taken as the assistant \textbf{\emph{A}}
(qubit 2). The spin-spin coupling constant is $J_{12}=214.95Hz$.
The relaxation times were $T_{1}=16.5$ s and $T_{2}=6.9$ s
for the proton, and $T_{1}=21.2$ s and $T_{2}=0.35$ s for the
carbon nuclei.

\subsection{Experimental procedure}

There are three steps to implement the measurement scheme stated
above: (\emph{i}) Preparation of the initial system, (\emph{ii})
Quantum simulation and (\emph{iii}) Measurement.

Any qubit state
$\hat{\rho}=\frac{1}{2}\mathbf{1}+\vec{s}\cdot\vec{\hat{S}}$ can be
parameterized as a vector in the Bloch sphere:
\begin{eqnarray}
\vec{s}(r,\theta,\phi)&=&(s_x,s_y,s_z)^T   \nonumber \\
&=&\left(r\sin\theta\cos\phi,r\sin\theta\cos\phi,r\cos\theta\right)^{T} ,
\label{state_g}
\end{eqnarray}
where the amplitude $r=1$ for a pure state, and $0\leq r<1$ for mixed
states and $\theta$, and $\phi$ are, respectively, the polar and
azimuthal angles.

The combined system was initialized in the state
$\hat{\varrho}_{0}=\hat{\rho}^{\left(S\right)}\otimes
\hat{\xi}^{\left(A\right)}$. In our demonstration experiment we
chose a completely disordered state
$\hat{\xi}=\frac{1}{2}\mathbf{1}^{\left(A\right)}$ that is
experimentally easy to prepare. Such an initial state
$\hat{\varrho}_{0}=\hat{\rho}^{\left(S\right)}\otimes
\frac{1}{2}\mathbf{1}^{\left(A\right)}$ was prepared by the NMR
pulse sequence:
\begin{equation}
\left[\arccos\left(r\right)\right]_{y}^{1}
\left[\frac{\pi}{2}\right]_{y}^{2}-G_{z}-\left[\theta\right]_{\phi + \pi/2}^{1} .
\end{equation}
The first two RF pulses define the amount of spin polarization on the two qubits.
The field gradient pulse $G_{z}$ dephases transverse magnetization to eliminate
off-diagonal terms in the density operator.
The last RF pulse turns the remaining (longitudinal) magnetization 
of qubit 1 into the desired orientation. 
The result of the preparation was checked using the standard method of state determination based on three noncommutative measurements of the system  \textbf{\emph{S}} , i.e., $\sigma_x^{(S)}$, $\sigma_y^{(S)}$ and $\sigma_z^{(S)}$. The experimental results are plotted in Fig. \ref{results} c) and d) for $r =1$ and $\frac{1}{2}$. The experimental average fidelity is above 0.99.

For the coupling Hamiltonian that transfers the information from
$S$ to $A$, we chose $\hat{H}^{opt}$ of example (c) (section IV B).
We performed two different methods for the simulation of this propagator  (\ref{e.U}):
the``short period expansion" (see section VI A), using the sequence
(\ref{Unmrde}) with $m=2$, and the exact
decomposition (\ref{e.Udecom})  (see section VI B) with the NMR
pulse sequences (\ref{e.pulseR}) and (\ref{e.pulseUd}).

After the coupling evolution, we measured the
\emph{x} components of the two spins to obtain the joint
probabilities $P_{kq}$.
For this purpose, we rotated the spins to the $z$-axis, using
a $\left[\frac{\pi}{2}\right]_{y}^{1,2}$ pulse and destroyed off-diagonal
elements by a magnetic field gradient pulse $G_z$.
The populations could then be measured by applying another rf pulse
to each of the spins and measuring their free induction decays (FIDs).
If the two spins are different isotopes, as in our case,
their FIDs usually have to be measured in separate experiments.

The resulting pulse sequence for the readout is thus
\begin{equation}
\left[\frac{\pi}{2}\right]_{-y}^{1,2}-G_{z}-\left[\frac{\pi}{2}\right]_{y}^{i}-FID_{i}.
\end {equation}
where $i=1$ or $2$ denotes qubit $i$. The measured $FIDs$
along with the normalization condition ($\sum_{kq}P_{kq}=1$) allowed
us to reconstruct the four diagonal elements (populations) in the
density matrix, which correspond to four joint probabilities
$P_{kq}$.
The information about the state $\hat{\rho}$
was then obtained by the inverse mapping $\mathcal{M}^{-1}$.

\subsection{Experimental Results}

\begin{figure}[tbh]
\begin{centering}
\includegraphics[width=\columnwidth]{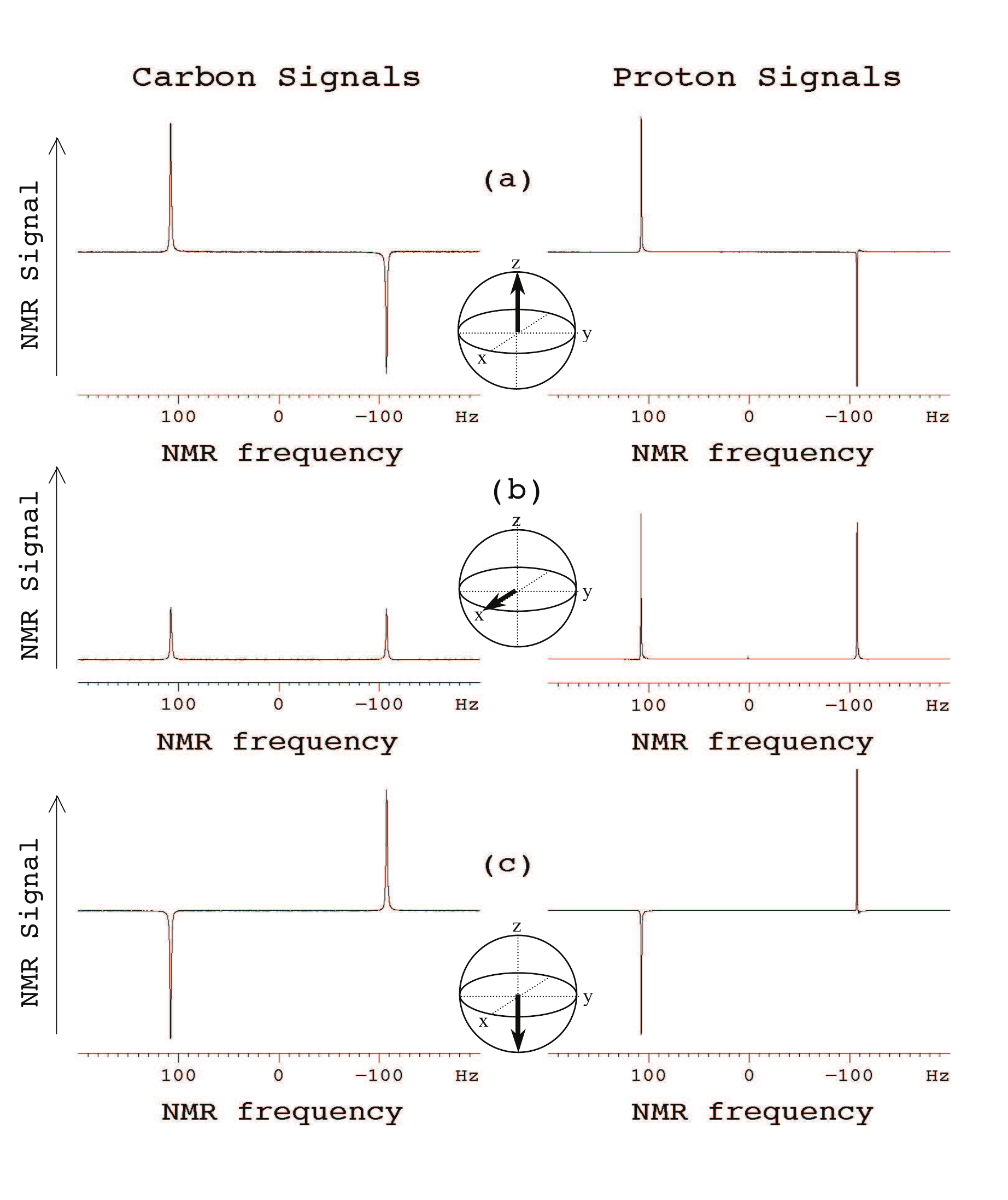}
\end{centering}
\caption{Experimental NMR spectra of carbon and proton for different initial conditions.
$\phi=0$ and 
(a)  $\theta=0$,  (b) $\theta=\pi/2$ and (c) $\theta=\pi$.  
The y-axis denotes the signal amplitude in arbitrary units.} 
\label{spectra}
\end{figure}

Figure \ref{spectra} shows the experimentally observed NMR signals after Fourier
transformation of the corresponding FIDs for the proton and carbon
spins for the following initial states:
(a) $\theta=0$ (i.e. $\hat{\rho}=\frac{1}{2}\mathbf{1}+\hat{S_{z}}$),
(b) $\phi=0, \theta=\pi/2$ (i.e., $\hat{\rho}=\frac{1}{2}\mathbf{1}+\hat{S_{x}}$),
(c) $\theta=\pi$ (i.e., $\hat{\rho}=\frac{1}{2}\mathbf{1}-\hat{S_{z}}$).

The amplitudes of the different resonance lines correspond directly to
population differences:
\begin{equation}
\begin{array}{c}
S_{NMR}(\textrm{Proton})\sim P_{1\mu}-P_{2\mu}\\
S_{NMR}(\textrm{Carbon})\sim P_{\mu1}-P_{\mu2} ,
\end{array}
\label{P_eq}
\end{equation}
 where $\mu=1$ for the resonance line with positive frequency and $\mu=2$ for
the negative frequency line.
From these populations, we determine the initial condition by
inverting Eq. (\ref{mapM}).

\begin{figure}[tbh]
\begin{centering}
\includegraphics[width=\columnwidth]{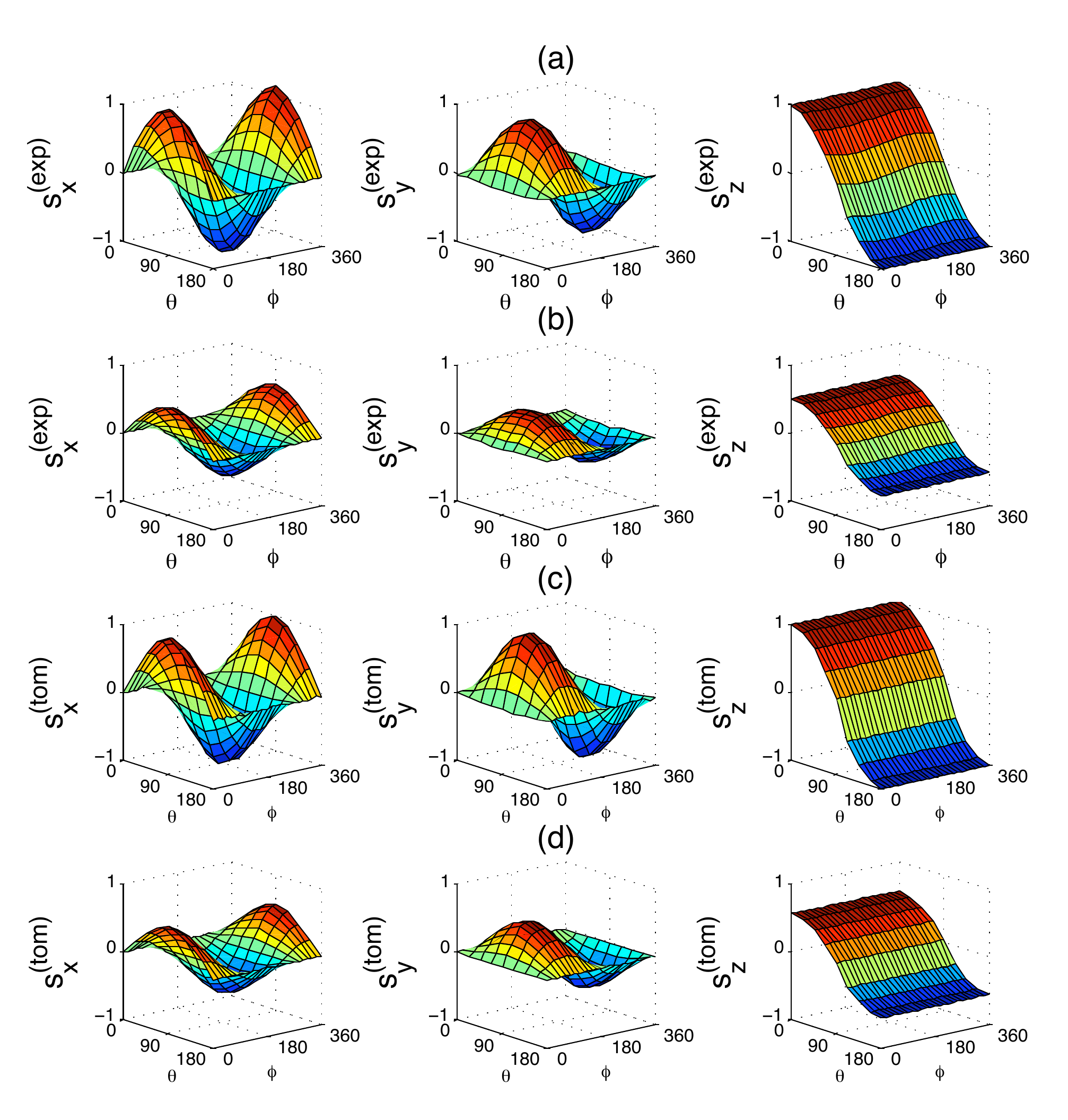}
\end{centering}
\caption{(Color Online) Experimental quantum state tomography for the general initial state 
$\vec{s}(r,\theta,\phi)$ [see Eq. (\ref{state_g})].
We compare the results from measuring a single observable of the combined system
\textit{\bf{S+A}} (rows (a) and (b)) with the results from the conventional measurement scheme
using three noncommutative measurements of the system \textit{\bf{S}} (rows (c) and (d)).
In both cases, the expectation values $s_{x},s_{y},s_{z}$, are shown from left to right
as functions of the angles $\theta$ and $\phi$: (a) and (c) for a pure state ($r=1$) and
(b) and (d) for a partially mixed state with $r=\frac{1}{2}$.}
\label{results}
\end{figure}

Fig. \ref{results} summarizes these results for a series of similar experiments,
where we chose initial conditions $\vec{s}\left(r,\theta,\phi\right)$
varying $\theta$ from 0 to $\pi$ in
increments of $\pi/8$, and $\phi$ from 0 to $2\pi$ with an increment of
$\pi/12$.
In Fig. \ref{results}a, we show the measured components
$s_{x}^{\left(\exp\right)}, s_{y}^{\left(\exp\right)}, s_{z}^{\left(\exp\right)}$
for pure states
($r=1$), while (b) shows the corresponding results for
mixed states with $r=\frac{1}{2}$.
The experiments cover a wide range of points on and within
the Bloch sphere.
The
experimental results clearly show the expected cosine and sine
modulations (\ref{state_g}), indicating that the measurement network is effective
for all these input states. The average fidelity over all
$N=9\times13$ measured states is
$$
F_{av}=\frac{1}{N}\sum_{1}^{N}
\frac{Tr(\hat{\rho}_{in}\hat{\rho}_{exp})}
{\sqrt{Tr(\hat{\rho}_{in}^{2})Tr(\hat{\rho}_{exp}^{2})}} \approx0.99
$$
for both cases, $r=1$ and $r=\frac{1}{2}$.

The experimental data shown in Fig. \ref{results} were obtained with
the ``short period expansion" technique of section VI A, i.e., the propagator $\hat{U}(\tau)$ was approximately realized by Eq. (\ref{Imp_Udt}) by repeating the sequence (\ref{Unmrde}) twice.
We also repeated the experiment with the ``exact decomposition" technique. The propagator $\hat{U}(\tau)$ was realized by the exact decomposition Eq. (\ref{e.Udecom}). The corresponding pulse sequence was obtained by combining the sequences (\ref{e.pulseR}) with (\ref{e.pulseUd}).  
The results that we obtained were similar to those represented in Fig. \ref{results},
but the fidelities were slightly lower.
This difference is probably due to the larger number of pulses in this experiment.

\subsection{Precision of the Measurement}

An alternative measure of the precision of the measurement
is the distance $D$ between the experimentally determined state
$\vec{s}_{exp}$ and the "true" input state $\vec{s}$.
In terms of the parametrization (\ref{state_g}), the trace distance
between the two density operators is
\begin{equation}
D(\vec{s}_{exp},\vec{s})=\frac{\left\vert
\vec{s}_{exp}-\vec{s}\right\vert }{2}=\frac{1}{2}\sqrt{\Delta
s_{x}^{2}+\Delta s_{y}^{2}+\Delta s_{z}^{2}}
\end{equation}
where $\Delta s_{\nu}=s_{\nu}^{(exp)}-s _{\nu}(\nu=x,y,z)$.

Writing $\Delta P$ for the experimental errors and using the definition
(\ref{mapM}) of the transfer matrix, we find for the distance
\begin{equation}
D(\vec{s}_{exp},\vec{s})=E|\Delta P|,
\end{equation}
where
 \begin{equation}
E=\frac{1}{2}\sqrt{E_{x}^{2}+E_{y}^{2}+E_{z}^{2}}
\label{C}
\end{equation}
and
\begin{equation}
\begin{array}{l}
E_{x}=\frac{\Delta s_{x}}{\Delta P}=\frac{1}{det(\tilde{\mathcal{M}})}\sum_{k=1}^{4} A_{k2} \\
E_{y}=\frac{\Delta s_{x}}{\Delta P}=\frac{1}{det(\tilde{\mathcal{M}})}\sum_{k=1}^{4} A_{k3} \\
E_{z}=\frac{\Delta s_{x}}{\Delta
P}=\frac{1}{det(\tilde{\mathcal{M}})}\sum_{k=1}^{4} A_{k4}
\end{array}.
\label{eQ_E}
\end{equation}
The $A_{kj}$ are the cofactors of the minors $\tilde{\mathcal{M}}_{kj}$ of
the transfer matrix 
$$
\tilde{\mathcal{M}} = \frac{1}{2} \mathcal{M}\left(
\begin{array}{cccc}
  1 & 0 & 0 & 1 \\
  0 & 1 & -i & 0 \\
  0 & 1 & i & 0 \\
  1 & 0 & 0 & -1 \\ 
\end{array}
\right) .
$$
The determinant of this matrix is $det(\tilde{\mathcal{M}})=-\frac{1}{4} i\Delta$.
Therefore, the error propagation coefficients $E_\alpha$ depend
only on the mapping $\mathcal{M}$. The smaller they are, the
higher the precision of the resulting measurement.

\begin{figure}[tbh]
\begin{center}
\includegraphics[width=0.98\columnwidth]{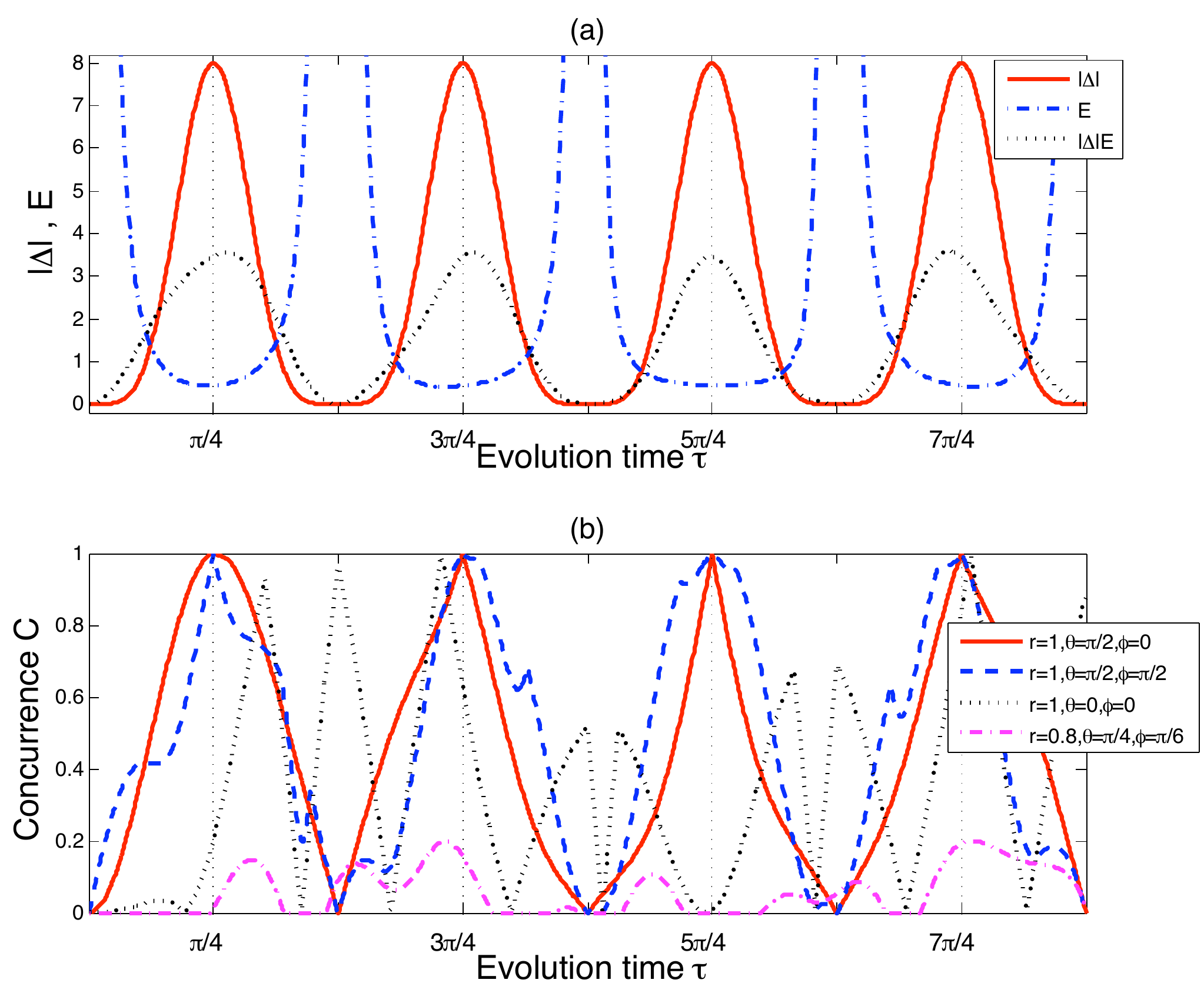}
\caption{(Color Online) (a) Error coefficient $E$ (solid line), magnitude of the
determinant $|\Delta|$ (dashed line) and their product (dotted line) vs. evolution time $\tau$ under the
Hamiltonian $\hat{H}^{opt}$ of the example (c) in Sec. IV B. (b)
The concurrence $C$ (i.e, entanglement) evolves with the evolution time $\tau$ under the
same Hamiltonian for the different initial states
$\hat{\rho}=\frac{1}{2} ( \mathbf{1} + r\sin\theta\cos\phi
\hat{S_x}+r\sin\theta\cos\phi\hat{S_y}+r\cos\theta\hat{S_z}) $ with
pure states of $r=1,\theta=\pi/2,\phi=0$ (red solid line),
$r=1,\theta=\pi/2,\phi=\pi/2$ (blue dashed line),
$r=1,\theta=0,\phi=0$ (black dotted line), and a specific mixed
state of $r=0.8,\theta=\pi/4,\phi=\pi/6$ (red dash-dotted line)}.
\label{AD}
\end{center}
\end{figure}

As Eq. (\ref{eQ_E}) shows, the error propagation scales inversely with the
determinant $\Delta$ of the transfer matrix $\mathcal{M}$.
We illustrate this dependence in Fig. \ref{AD}a), where we plot the two
quantities as a function of the coupling evolution time $\tau$.
The minima of $E$
occur near the maxima of $|\Delta|$, and when $|\Delta|=0$, $E$ tends
to infinity.
In this range, it is impossible to determine the state $\rho$
by such a measurement.
A closer look shows that the minima of $E$ do
not occur exactly at the maxima of $|\Delta|$.
The difference arises from the numerators in (\ref{eQ_E}).

In our experiment, the experimental uncertainties are
$\Delta P \approx 5 \%$.
For the chosen experimental parameters, this results in an
average distance $D_{av}(\vec{s}_{exp},\vec{s})=0.04$ for $r=1$
and $0.03$ for $r=\frac{1}{2}$.
The distance measurement $D$ and the fidelity measurement F are
related by $1-F\leq D\leq\sqrt{1-F^{2}}$ \cite{NielsenBook:2000aa}.

\subsection{Entanglement}

The evolution that transfers information from the system to the
assistant can entangle the two qubits with each other.
In Fig.  \ref{AD}, we quantify the entanglement generated
and relate it to the precision of the measurement.
Fig. \ref{AD} b)  shows the concurrence $C$ during the coupling
evolution, calculated as
$$
C(t)=max\{\chi_{1}-\chi_{2}-\chi_{3}-\chi_{4},0\},
$$
where $\chi_{i}(i=1,2,3,4)$ are the square roots of the eigenvalues
of
$$
16 \, \hat{\varrho}_{t}(S_{y}^{1}S_{y}^{2})\hat{\varrho}_{t}^{*}(S_{y}^{1}S_{y}^{2})
$$
in decreasing order, and
$$
\hat{\varrho}_{t}=\hat{U}(t)(\hat{\rho}\otimes\frac{1}{2}\mathbf{1})\hat{U}^{\dagger}(t)
$$
is the instantaneous density operator.

If the initial state is in the $xy$-plane, the entanglement between the system
and assistant is maximized at roughly the same time as the information transfer
for these measurements is optimized (as quantified by $|\Delta|$).
However, for initial conditions oriented along the z-axis, the entanglement
generated by the specific Hamiltonian shows a relatively
complicated time dependence and little correlation with the precision of the measurement.
For evolution times close to $\tau \approx 5 \pi /4$, e.g., the 
entanglement vanishes, while the measurement error is minimized.

The dash-dotted curve in Fig. \ref{AD}b) shows the entanglement that is generated
for a partially mixed input state with a general orientation ($r = 0.8, \theta = \pi/4, \phi = \pi/6$).
In this case,  the concurrence remains below 0.2 and reaches zero even at the
times where the measurement precision is optimized.
If the amplitude $r$ is reduced further, the entanglement vanishes,
$C(\hat{\varrho}_{t})\equiv0$, but the precision of the measurement is not affected.
We conclude that entanglement between system and assistant is not
an essential criterion for the success of this measurement scheme.

\section{CONCLUSION}

We have experimentally demonstrated how the complete state of a quantum
system can be obtained from the results of 
repeated measurements with a single, factorized observable $\hat{\Omega}$.
The procedure, which involves a controlled interaction between the
system under test and a second quantum system, was proposed
by Allahverdyan et al. \cite{Allahverdyan:2004aa}.

In our experiment, we used a Heisenberg-coupling to transfer information
from the system to the assistant.
Interactions of this type are found in many physical systems:
apart from nuclear spins (like in this work), they also occur in
quantum dots \cite{Loss:1998aa,Imamogu:1999aa}, donor atoms in silicon \cite{Kane:1998aa, Vrijen:2000aa},
quantum Hall systems \cite{Mozyrsky:2001aa} and electrons on
helium \cite{Dykman:2000aa}.

The precision of this type of measurements depends strongly on the
details of the interaction between system and assistant,
on the type of Hamiltonian as well as on the duration of the interaction.
This can be understood by considering the transfer of information from
the state of the system to the measurement results from the single observable:
If we describe this transfer of information from $n$ elements of the density
operator of the input state by a matrix $\mathcal{M}$, the rank of this
matrix must be $n$, i.e. its inverse must exist.
In practice, it is necessary to choose a transfer matrix that is far from
the singular case, to maximize the precision with which the input state
can be calculated from the measurement results.

This initial work has demonstrated the basic possibility of implementing such
measurements on the simplest possible quantum system (a single spin 1/2).
Of course it is possible to extend the scheme to systems of arbitrary size.
Work in this direction is currently under way.

\begin{center}\textbf{ACKNOWLEDGMENTS} \par\end{center}

We gratefully acknowledge helpful discussions with Dr. Bo Chong, Dr. Jingfu Zhang
and financial support from the DFG through Su 192/19-1.  Du. J thanks the support of NSFC of China, CAS and the European Commission under Contract No. 007065 (Marie Curie).


\begin{thebibliography}{26}
\expandafter\ifx\csname natexlab\endcsname\relax\def\natexlab#1{#1}\fi
\expandafter\ifx\csname bibnamefont\endcsname\relax
  \def\bibnamefont#1{#1}\fi
\expandafter\ifx\csname bibfnamefont\endcsname\relax
  \def\bibfnamefont#1{#1}\fi
\expandafter\ifx\csname citenamefont\endcsname\relax
  \def\citenamefont#1{#1}\fi
\expandafter\ifx\csname url\endcsname\relax
  \def\url#1{\texttt{#1}}\fi
\expandafter\ifx\csname urlprefix\endcsname\relax\def\urlprefix{URL }\fi
\providecommand{\bibinfo}[2]{#2}
\providecommand{\eprint}[2][]{\url{#2}}

\bibitem[{\citenamefont{Pauli}(1933)}]{Pauli:1933aa}
\bibinfo{author}{\bibfnamefont{W.}~\bibnamefont{Pauli}}, in
  \emph{\bibinfo{booktitle}{Handbuch der Physik}}, edited by
  \bibinfo{editor}{\bibfnamefont{H.}~\bibnamefont{Geiger}} \bibnamefont{and}
  \bibinfo{editor}{\bibfnamefont{K.}~\bibnamefont{Scheel}}
  (\bibinfo{publisher}{Springer}, \bibinfo{address}{Berlin},
  \bibinfo{year}{1933}), vol.~\bibinfo{volume}{24}, p.~\bibinfo{pages}{98}.

\bibitem[{\citenamefont{Bohr}(1935)}]{Bohr:1935aa}
\bibinfo{author}{\bibfnamefont{N.}~\bibnamefont{Bohr}}, \bibinfo{journal}{Phys.
  Rev.} \textbf{\bibinfo{volume}{48}}, \bibinfo{pages}{696}
  (\bibinfo{year}{1935}).

\bibitem[{\citenamefont{Helstrom}(1976)}]{HelstromBook:1976aa}
\bibinfo{author}{\bibfnamefont{C.~W.} \bibnamefont{Helstrom}},
  \emph{\bibinfo{title}{Quantum Detection and Estimation Theory}}
  (\bibinfo{publisher}{Academic Press}, \bibinfo{address}{New York},
  \bibinfo{year}{1976}).

\bibitem[{\citenamefont{Chefles}(2000)}]{Chefles:2000aa}
\bibinfo{author}{\bibfnamefont{A.}~\bibnamefont{Chefles}},
  \bibinfo{journal}{Contemporary Physics} \textbf{\bibinfo{volume}{41}},
  \bibinfo{pages}{401} (\bibinfo{year}{2000}).

\bibitem[{\citenamefont{Gill and Guta}(2003)}]{Gill:2003aa}
\bibinfo{author}{\bibfnamefont{R.}~\bibnamefont{Gill}} \bibnamefont{and}
  \bibinfo{author}{\bibfnamefont{M.}~\bibnamefont{Guta}},
  \bibinfo{journal}{arXiv.org:quant-ph/0303020}  (\bibinfo{year}{2003}).

\bibitem[{\citenamefont{Bechmann-Pasquinucci and
  Tittel}(2000)}]{Bechmann-Pasquinucci:2000aa}
\bibinfo{author}{\bibfnamefont{H.}~\bibnamefont{Bechmann-Pasquinucci}}
  \bibnamefont{and} \bibinfo{author}{\bibfnamefont{W.}~\bibnamefont{Tittel}},
  \bibinfo{journal}{Phys. Rev. A} \textbf{\bibinfo{volume}{61}},
  \bibinfo{pages}{062308} (\bibinfo{year}{2000}).

\bibitem[{\citenamefont{Bechmann-Pasquinucci and
  Peres}(2000)}]{Bechmann-Pasquinucci:2000ab}
\bibinfo{author}{\bibfnamefont{H.}~\bibnamefont{Bechmann-Pasquinucci}}
  \bibnamefont{and} \bibinfo{author}{\bibfnamefont{A.}~\bibnamefont{Peres}},
  \bibinfo{journal}{Phys. Rev. Lett.} \textbf{\bibinfo{volume}{85}},
  \bibinfo{pages}{3313} (\bibinfo{year}{2000}).

\bibitem[{\citenamefont{Welsch et~al.}(1999)\citenamefont{Welsch, Vogel, and
  Opatrn{\'y}}}]{Welsch:1999aa}
\bibinfo{author}{\bibfnamefont{D.}~\bibnamefont{Welsch}},
  \bibinfo{author}{\bibfnamefont{W.}~\bibnamefont{Vogel}}, \bibnamefont{and}
  \bibinfo{author}{\bibfnamefont{T.}~\bibnamefont{Opatrn{\'y}}},
  \bibinfo{journal}{Progr. Opt.} \textbf{\bibinfo{volume}{39}},
  \bibinfo{pages}{63} (\bibinfo{year}{1999}).

\bibitem[{\citenamefont{D'Ariano}(1997)}]{DAriano:1997aa}
\bibinfo{author}{\bibfnamefont{G.}~\bibnamefont{D'Ariano}},
  \bibinfo{journal}{In: T. HakioImagelu and A.S. Shumovsky, Editors, Quantum
  Optics and Spectroscopy of Solids, Editors, Kluwer, Amsterdam} pp.
  \bibinfo{pages}{175--202} (\bibinfo{year}{1997}).

\bibitem[{\citenamefont{Leonhardt}(1997)}]{LeonhardtBook:1997aa}
\bibinfo{author}{\bibfnamefont{U.}~\bibnamefont{Leonhardt}},
  \emph{\bibinfo{title}{Measuring the Quantum State of Light}}
  (\bibinfo{publisher}{Cambridge University Press}, \bibinfo{address}{New
  York}, \bibinfo{year}{1997}).

\bibitem[{\citenamefont{Ivonovic}(1981)}]{Ivonovic:1981aa}
\bibinfo{author}{\bibfnamefont{I.~D.} \bibnamefont{Ivonovic}},
  \bibinfo{journal}{J. Phys. A: Math. Gen.} \textbf{\bibinfo{volume}{14}},
  \bibinfo{pages}{3241} (\bibinfo{year}{1981}).

\bibitem[{\citenamefont{Wootters and Fields}(1989)}]{Wootters:1989aa}
\bibinfo{author}{\bibfnamefont{W.~K.} \bibnamefont{Wootters}} \bibnamefont{and}
  \bibinfo{author}{\bibfnamefont{B.~D.} \bibnamefont{Fields}},
  \bibinfo{journal}{Ann. Phys.} \textbf{\bibinfo{volume}{191}},
  \bibinfo{pages}{363} (\bibinfo{year}{1989}).

\bibitem[{\citenamefont{Rehacek et~al.}(2004)\citenamefont{Rehacek, Englert,
  and Kaszlikowski}}]{Rehacek:2004aa}
\bibinfo{author}{\bibfnamefont{J.}~\bibnamefont{Rehacek}},
  \bibinfo{author}{\bibfnamefont{B.-G.} \bibnamefont{Englert}},
  \bibnamefont{and}
  \bibinfo{author}{\bibfnamefont{D.}~\bibnamefont{Kaszlikowski}},
  \bibinfo{journal}{Phys. Rev. A} \textbf{\bibinfo{volume}{70}},
  \bibinfo{eid}{052321} (\bibinfo{year}{2004}).

\bibitem[{\citenamefont{D'Ariano}(2002)}]{DAriano:2002aa}
\bibinfo{author}{\bibfnamefont{G.~M.} \bibnamefont{D'Ariano}},
  \bibinfo{journal}{Phys. Lett. A} \textbf{\bibinfo{volume}{300}},
  \bibinfo{pages}{1} (\bibinfo{year}{2002}).

\bibitem[{\citenamefont{Caves et~al.}(2002)\citenamefont{Caves, Fuchs, and
  Schack}}]{Caves:2002aa}
\bibinfo{author}{\bibfnamefont{C.~M.} \bibnamefont{Caves}},
  \bibinfo{author}{\bibfnamefont{C.~A.} \bibnamefont{Fuchs}}, \bibnamefont{and}
  \bibinfo{author}{\bibfnamefont{R.}~\bibnamefont{Schack}},
  \bibinfo{journal}{Journal of Mathematical Physics}
  \textbf{\bibinfo{volume}{43}}, \bibinfo{pages}{4537} (\bibinfo{year}{2002}),
  \urlprefix\url{http://link.aip.org/link/?JMP/43/4537/1}.

\bibitem[{\citenamefont{Du et~al.}(2006)\citenamefont{Du, Sun, Peng, and
  Durt}}]{Du:2006aa}
\bibinfo{author}{\bibfnamefont{J.}~\bibnamefont{Du}},
  \bibinfo{author}{\bibfnamefont{M.}~\bibnamefont{Sun}},
  \bibinfo{author}{\bibfnamefont{X.}~\bibnamefont{Peng}}, \bibnamefont{and}
  \bibinfo{author}{\bibfnamefont{T.}~\bibnamefont{Durt}},
  \bibinfo{journal}{Phys. Rev. A} \textbf{\bibinfo{volume}{74}},
  \bibinfo{eid}{042341} (\bibinfo{year}{2006}).

\bibitem[{\citenamefont{Allahverdyan et~al.}(2004)\citenamefont{Allahverdyan,
  Balian, and Nieuwenhuizen}}]{Allahverdyan:2004aa}
\bibinfo{author}{\bibfnamefont{A.~E.} \bibnamefont{Allahverdyan}},
  \bibinfo{author}{\bibfnamefont{R.}~\bibnamefont{Balian}}, \bibnamefont{and}
  \bibinfo{author}{\bibfnamefont{T.~M.} \bibnamefont{Nieuwenhuizen}},
  \bibinfo{journal}{Phys. Rev. Lett.} \textbf{\bibinfo{volume}{92}},
  \bibinfo{eid}{120402} (\bibinfo{year}{2004}).

\bibitem[{\citenamefont{Ernst et~al.}(1994)\citenamefont{Ernst, Bodenhausen,
  and Wokaun}}]{ErnstBook:1994aa}
\bibinfo{author}{\bibfnamefont{R.~R.} \bibnamefont{Ernst}},
  \bibinfo{author}{\bibfnamefont{G.}~\bibnamefont{Bodenhausen}},
  \bibnamefont{and} \bibinfo{author}{\bibfnamefont{A.}~\bibnamefont{Wokaun}},
  \emph{\bibinfo{title}{Principles of Nuclear Magnetic Resonance in One and Two
  Dimensions}} (\bibinfo{publisher}{Oxford University Press},
  \bibinfo{address}{Oxford}, \bibinfo{year}{1994}).

\bibitem[{\citenamefont{Trotter}(1959)}]{Trotter:1959aa}
\bibinfo{author}{\bibfnamefont{H.}~\bibnamefont{Trotter}},
  \bibinfo{journal}{Proc. Amer. Math. Soc.} \textbf{\bibinfo{volume}{10}},
  \bibinfo{pages}{545} (\bibinfo{year}{1959}).

\bibitem[{\citenamefont{Nielsen and Chuang}(2000)}]{NielsenBook:2000aa}
\bibinfo{author}{\bibfnamefont{M.~A.} \bibnamefont{Nielsen}} \bibnamefont{and}
  \bibinfo{author}{\bibfnamefont{I.~L.} \bibnamefont{Chuang}},
  \emph{\bibinfo{title}{Quantum Computation and Quantum Information}}
  (\bibinfo{publisher}{Cambridge Univ. Press}, \bibinfo{address}{Cambridge},
  \bibinfo{year}{2000}).

\bibitem[{\citenamefont{Loss and DiVincenzo}(1998)}]{Loss:1998aa}
\bibinfo{author}{\bibfnamefont{D.}~\bibnamefont{Loss}} \bibnamefont{and}
  \bibinfo{author}{\bibfnamefont{D.~P.} \bibnamefont{DiVincenzo}},
  \bibinfo{journal}{Phys. Rev. A} \textbf{\bibinfo{volume}{57}},
  \bibinfo{pages}{120} (\bibinfo{year}{1998}).

\bibitem[{\citenamefont{Imamogu et~al.}(1999)\citenamefont{Imamogu, Awschalom,
  Burkard, DiVincenzo, Loss, Sherwin, and Small}}]{Imamogu:1999aa}
\bibinfo{author}{\bibfnamefont{A.}~\bibnamefont{Imamogu}},
  \bibinfo{author}{\bibfnamefont{D.~D.} \bibnamefont{Awschalom}},
  \bibinfo{author}{\bibfnamefont{G.}~\bibnamefont{Burkard}},
  \bibinfo{author}{\bibfnamefont{D.~P.} \bibnamefont{DiVincenzo}},
  \bibinfo{author}{\bibfnamefont{D.}~\bibnamefont{Loss}},
  \bibinfo{author}{\bibfnamefont{M.}~\bibnamefont{Sherwin}}, \bibnamefont{and}
  \bibinfo{author}{\bibfnamefont{A.}~\bibnamefont{Small}},
  \bibinfo{journal}{Phys. Rev. Lett.} \textbf{\bibinfo{volume}{83}},
  \bibinfo{pages}{4204} (\bibinfo{year}{1999}).

\bibitem[{\citenamefont{Kane}(1998)}]{Kane:1998aa}
\bibinfo{author}{\bibfnamefont{B.~E.} \bibnamefont{Kane}},
  \bibinfo{journal}{Nature} \textbf{\bibinfo{volume}{393}},
  \bibinfo{pages}{133} (\bibinfo{year}{1998}).

\bibitem[{\citenamefont{Vrijen et~al.}(2000)\citenamefont{Vrijen, Yablonovitch,
  Wang, Jiang, Balandin, Roychowdhury, Mor, and DiVincenzo}}]{Vrijen:2000aa}
\bibinfo{author}{\bibfnamefont{R.}~\bibnamefont{Vrijen}},
  \bibinfo{author}{\bibfnamefont{E.}~\bibnamefont{Yablonovitch}},
  \bibinfo{author}{\bibfnamefont{K.}~\bibnamefont{Wang}},
  \bibinfo{author}{\bibfnamefont{H.~W.} \bibnamefont{Jiang}},
  \bibinfo{author}{\bibfnamefont{A.}~\bibnamefont{Balandin}},
  \bibinfo{author}{\bibfnamefont{V.}~\bibnamefont{Roychowdhury}},
  \bibinfo{author}{\bibfnamefont{T.}~\bibnamefont{Mor}}, \bibnamefont{and}
  \bibinfo{author}{\bibfnamefont{D.}~\bibnamefont{DiVincenzo}},
  \bibinfo{journal}{Phys. Rev. A} \textbf{\bibinfo{volume}{62}},
  \bibinfo{pages}{012306} (\bibinfo{year}{2000}).

\bibitem[{\citenamefont{Mozyrsky et~al.}(2001)\citenamefont{Mozyrsky, Privman,
  and Glasser}}]{Mozyrsky:2001aa}
\bibinfo{author}{\bibfnamefont{D.}~\bibnamefont{Mozyrsky}},
  \bibinfo{author}{\bibfnamefont{V.}~\bibnamefont{Privman}}, \bibnamefont{and}
  \bibinfo{author}{\bibfnamefont{M.~L.} \bibnamefont{Glasser}},
  \bibinfo{journal}{Phys. Rev. Lett.} \textbf{\bibinfo{volume}{86}},
  \bibinfo{pages}{5112} (\bibinfo{year}{2001}).

\bibitem[{\citenamefont{Dykman and Platzman}(2000)}]{Dykman:2000aa}
\bibinfo{author}{\bibfnamefont{M.}~\bibnamefont{Dykman}} \bibnamefont{and}
  \bibinfo{author}{\bibfnamefont{P.}~\bibnamefont{Platzman}},
  \bibinfo{journal}{Fortschritte der Physik} \textbf{\bibinfo{volume}{48}},
  \bibinfo{pages}{1095} (\bibinfo{year}{2000}).

\end{thebibliography}
\end{document}